\documentclass[aps,10pt,pra,twocolumn,showpacs,showkeys,groupedaddress]{revtex4-1}

\usepackage[pagebackref,colorlinks,linkcolor=red,citecolor=green,filecolor=cyan,urlcolor=magenta]{hyperref}	
\renewcommand*\backref[1]{\ifx#1\relax \else {\color{black}(cited on page~#1).} \fi}
\usepackage{amsmath}
\usepackage{graphicx}
\usepackage{bm}
\usepackage{xcolor}
\usepackage[top=14mm,bottom=20mm,left=20mm,right=20mm]{geometry}

\def\mum{{\rm\mu m}}

\def\muK{{\rm\mu K}}
\def\ie{{\it i.e.,\/}}

\usepackage[normalem]{ulem}

\newcommand{\be}{\begin{equation}}
\newcommand{\ee}{\end{equation}}
\newcommand{\bea}{\begin{eqnarray}}
\newcommand{\eea}{\end{eqnarray}}

\begin{document}
\title{Suppression and enhancement of decoherence in an atomic Josephson junction}

\author{Yonathan Japha, Shuyu Zhou, Mark Keil and Ron Folman}
\affiliation{Department of Physics, Ben-Gurion University
of the Negev,
Beer-Sheva 84105, Israel}
\author{Carsten Henkel}
\affiliation{Institute of Physics and Astronomy, University of Potsdam, 14476 Potsdam, Germany}
\author{Amichay Vardi}
\affiliation{Department of Chemistry,
Ben-Gurion University
of the Negev, Beer-Sheva 84105, Israel}

\begin{abstract}
We examine the role of interactions for a Bose gas trapped in a double-well potential (``Bose-Josephson junction'') when external noise is applied and the system is initially delocalized with equal probability amplitudes in both sites. The noise may have two kinds of effects: loss of atoms from the trap, and random shifts in the relative phase or number difference between the two wells. The effects of phase noise are mitigated by atom-atom interactions and tunneling, such that the dephasing rate may be reduced to half its single-atom value. Decoherence due to number noise (which induces fluctuations in the relative atom number between the wells) is considerably enhanced by the interactions. A similar scenario is predicted for the case of atom loss, even if the loss rates from the two sites are equal. In fact, interactions convert the increased uncertainty in atom number (difference) into (relative) phase diffusion and reduce the coherence across the junction. We examine the parameters relevant for these effects using a simple model of the trapping potential based on an atom chip device. These results provide a framework for mapping the dynamical range of barriers engineered for specific applications and sets the stage for more complex circuits (``atomtronics'').
\end{abstract}

\maketitle

\section{Introduction}

The development of circuits for neutral atoms (coined ``atomtronics'') has received tremendous impetus from recent advances in the control and manipulation of ultracold atoms using magnetic and optical fields ~\cite{Pepino,Charron,Japha}.
The idea of atomtronics is inspired by the analogy between ultracold atoms confined in optical or magnetic
potentials and solid-state systems based on electrons in various forms of conductors, semiconductors or superconductors.
For example, ultracold atoms in optical lattices exhibit a Mott insulator to superfluid transition, or display spin-orbit
coupling as in solid-state systems.
Another example is a Bose-Einstein condensate (BEC) of neutral atoms in a double-well potential, which is
analogous to a Josephson junction of coupled superconductors. On the other hand, the quantum
properties of ultracold atoms as coherent matter waves enable
systems that are equivalent to
optical circuits, which are based on waveguides and beam-splitters for interferometric precision measurements
in fundamental science and technological applications.

A promising platform for accurately manipulating matter waves in a way that would enable integrated
circuits for neutral atoms is an atom chip~\cite{ReichelBook,Reichel2002,Folman2002,Zimmermann2007}.
Such a device  facilitates precise control over magnetic or optical potentials on the micrometer scale. This length
scale, which is on the order of the de-Broglie wavelength of ultracold atoms under typical conditions, permits
control of important dynamical parameters, such as the tunneling rate through a potential barrier.
For a network built of static magnetic fields, such control over the dynamics requires loading the atoms into
potentials just a few $\mu$m from the surface of the chip~\cite{Salem}.
The ability to load such potentials, while maintaining spatial coherence, was recently shown to be possible~\cite{Shuyu}.
This achievement is facilitated by the weak coupling of neutral atoms to the environment~\cite{Carsten2003}.
Yet, in view of the fact that spatial coherence is one of the most vulnerable properties of quantum systems made
of massive particles, it is quite surprising that a BEC of thousands of atoms preserves spatial coherence for
a relatively long time in the very close proximity of a few micrometers from a conducting surface at room temperature.

Here we examine the interplay between coupling to external noise and the internal parameters -- tunneling rate
and atom-atom interactions -- of a BEC in a double-well potential (a ``Bose-Josephson junction'').
Such a system, consisting of a potential barrier between two potential wells, provides one of the fundamental
building blocks of atomic circuits and comprises one of the basic models for studying a 
simple system of many interacting particles occupying only two modes.
This study unravels some general many-body effects,
and at the same time enables insights into the limits of
the practical use of circuits of a trapped BEC near an atom chip surface.

Macroscopic one-particle coherence is the hallmark of Bose-Einstein condensation.
The Penrose-Onsager criterion states that as the condensate forms, one of the eigenvalues of the reduced
one-particle density matrix becomes dominant, resulting in a pure state in which all atoms occupy the
same quantum ``orbital''.
Once a condensate is prepared however, its one-particle coherence can be lost via entanglement with an external environment (``decoherence'') or by internal entanglement between condensate atoms due to interactions (``phase diffusion''). The interplay between these two processes, namely non-Hamiltonian decoherence and the Hamiltonian dynamics of interacting particles, is often rich and intricate. The combined effect is rarely additive and depends strongly on the details of the coupling mechanisms. For example, decoherence may be used to {\em protect} one-particle coherence by suppressing interaction-induced squeezing in a quantum-Zeno-like effect~\cite{Khodorkovsky2008,Khodorkovsky2009}.
It may also induce stochastic resonances which enhance the system's response to external driving~\cite{Witthaut2009}.

Reversing roles, one may ask how interactions affect the dephasing or dissipation of a BEC due to its coupling to the environment. In this work we consider a BEC in a double well using a two-mode Bose-Hubbard approximation, and investigate how the loss of one-particle coherence due to the external noise is affected by interparticle interactions. We find that many-body dynamics may indeed either enhance or suppress decoherence, depending on the nature of the applied noise.

In light of these fundamental effects, this work also attempts
to construct a framework for combining our theoretical model with practical experimental parameters for realistic magnetic potentials
and magnetic noise on an atom chip. Finally, we present some experimental results concerning
atom loss at distances of a few micrometers from an atom chip, which may enable some concrete conclusions
regarding the issue of coherence in atomic circuits using similar platforms.

This paper is structured as follows:
in Sec.~\ref{sec:system}, we describe the basic constituents
of the system we are about to study.
In Sec.~\ref{sec:TMBH} we review the theoretical model and
fundamental properties of a BEC in a double well and in Sec.~\ref{sec:magnoise} we derive its coupling to magnetic noise.
Section~\ref{sec:main} then combines these effects to present the main results of this work:
how decoherence in an atomic Josephson junction can be suppressed or enhanced by atom-atom interactions.
The range of validity and accessible range of parameters of our
model are discussed in Sec.~\ref{sec:validity}.
This discussion is supplemented by experimental measurements of magnetic noise in atomic traps (Sec.~\ref{sec:lifetime}).
Finally, our discussion in Sec.~\ref{sec:discussion} includes examples of practical and fundamental implications of the
predicted effects.

\section{Description of the system}
\label{sec:system}

We consider a Bose-Einstein condensate (BEC) of atoms with mass $m$ in a double-well potential.
The potential is modeled by a cylindrically symmetric harmonic
transverse part $V_{\perp}=\frac{1}{2}m\omega_{\perp}^2[y^2+(z-z_0)^2]$
centered at a distance $z_0$ from the surface of an atom chip,
and a longitudinal part $V_{\parallel}$ representing a barrier of height $V_0$ between two wells with minima
at $x=\pm d/2$. This potential, which is symmetric under $x\to -x$ reflections, may be modeled as:
\be
V_{\parallel}(x)=\left\{\begin{array}{lr} V_0\cos^2(\pi x/d) & |x|\leq d/2 \\ \frac{1}{2}m\omega_x^2(|x|-d/2)^2 & |x|>d/2
\end{array}\right. \, ,
\label{eq:Vxyz}
\ee
where the longitudinal frequency $\omega_x$ characterizing the curvature of the potential far from the barrier
is typically smaller than the transverse frequency $\omega_{\perp}$.
Such a system is usually referred to as a Josephson junction; it exhibits Josephson oscillations with frequency $\omega_J$
when the number of atoms in the two wells deviates slightly from equilibrium.

Some of the most important properties of the condensate in the potential can be derived from the assumption that a ``macroscopic" number of atoms occupy a single mode whose wave function $\phi_0({\bf r})$
satisfies the Gross-Pitaevskii (GP) equation, whose static form is
\be [-(\hbar^2/2m)\nabla^2+V({\bf r})+gN|\phi_0({\bf r})|^2-\mu]\phi_0({\bf r})=0,
\label{eq:GP-ground-state}
\ee
where $\mu$ is the chemical potential, which is the energy of a single atom in the effective (mean-field) potential $V_{\rm eff}=V({\bf r})+gN|\phi_0({\bf r})|^2$. 
Here $N$ is the total number of atoms
and $g=4\pi\hbar^2 a_s/m$ is the collisional interaction strength, with $a_s$ being the $s$-wave scattering length.

As long as the barrier height is not too high and the temperature is low enough, the atoms predominantly occupy the
condensate mode $\phi_0({\bf r})$ and the system is coherent, namely, the phase
between the two sites is well defined. However, when the barrier height grows and tunneling is suppressed,
more atoms occupy other spatial modes and the one-particle coherence drops.
In order to understand this effect, we use a set of spatial modes defined by higher-energy solutions of
the GP equation
\be [-(\hbar^2/2m)\nabla^2+V({\bf r})+gN|\phi_0({\bf r})|^2-\mu]\phi_j({\bf r})=E_j\phi_j({\bf r}),
\label{eq:GP-excited-states}
\ee
where $\phi_0({\bf r})$ is the condensate mode with $E_0=0$ and the other modes ($j>0$) represent excited
single-atom states in the mean-field potential~\cite{JaphaBand2011}. These modes form a complete set which may serve as a basis for any calculation.
Furthermore, this specific choice is useful because it can describe the ground state of the system for all barrier
heights. 

\begin{figure}[h!]
      \centering
      \includegraphics[width=\columnwidth]{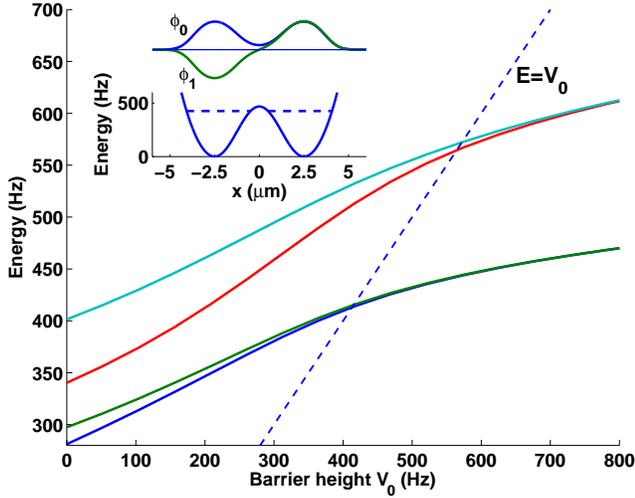} \\
      \caption{Mean-field energy levels of the lowest energy spatial modes for $N=200$ $^{87}$Rb atoms
as a function of barrier height $V_0$
in a double-well potential with transverse frequency $\omega_{\perp}=2\pi\times 500\,{\rm Hz}$ and a longitudinal
potential [Eq.~(\protect\ref{eq:Vxyz})] with $d=5\,\mu$m and $\omega_x=2\pi\times 200\,{\rm Hz}$.
The lowest solid curve represents the longitudinal energy $\mu_{\parallel}$
of the condensate
[Eq.~(\ref{eq:def-H-parallel})],
and the other energies are higher than
this energy by $E_j$ for $j=1,2,3$
[Eq.~(\ref{eq:GP-excited-states})]. When the barrier height grows, the energy splitting between the
lowest energy pair decreases and the anti-symmetric mode $\phi_1$ becomes
significantly occupied even at very low temperatures or even at absolute zero due to mixing between the
two lowest levels caused by atom-atom interactions.
The inset shows the shape of the potential for a barrier height
$V_0\approx 470\,{\rm Hz}$
where $\mu_{\parallel}\approx 425\,{\rm Hz}$
 (dashed line) is slightly lower than the barrier.
The symmetric condensate wavefunction $\phi_0=(\phi_R+\phi_L)/\sqrt{2}$ and the antisymmetric wavefunction
$\phi_1=(\phi_R-\phi_L)/\sqrt{2}$  have approximately the same shape in the two wells, and
the energy splitting is $E_1\approx 1\,{\rm Hz}$.}
\label{fig:Elevels}
\end{figure}

When the barrier is low (or does not exist) the condensate approximation holds for the ground state.
However, the nature of the ground state changes when the barrier becomes higher than the longitudinal ground-state
energy
\begin{eqnarray}
\mu_{\parallel} &\equiv&  \int d^3 r\ \phi_0({\bf r})\hat{H}_{\parallel}^{\rm eff}\phi_0({\bf r})\, ; \nonumber \\
\hat{H}_{\parallel}^{\rm eff} &\equiv & -\frac{\hbar^2}{2m}\frac{\partial^2}{\partial x^2}+
V_{\parallel}(x)+gN|\phi_0(x,y=0,z=z_0)|^2.\quad\
	\label{eq:def-H-parallel}
\end{eqnarray}
As demonstrated in Fig.~\ref{fig:Elevels}, when $\mu_{\parallel}\lesssim V_0$
the energy $E_1$ of the (anti-symmetric) first excited mode $\phi_1$ becomes very small
compared to the other excited modes and the configuration space can be described by
the two modes $\phi_0$ and $\phi_1$, or alternatively by their superpositions $\phi_L$ and $\phi_R$
localized predominantly in the left- and right-hand wells, respectively.
In this regime, collisional interactions
play a major role in determining the ground state configuration beyond the spatial effects accounted for
by the mean-field potential, as will be described in Sec.~\ref{sec:TMBH}.

In this work we focus on the interplay between the intrinsic parameters of the system and
the coupling to the environment. The latter may appear in different shapes and forms. Basically, the strongest coupling is between the magnetic moment of the atom
and magnetic field noise originating from current fluctuations in the atom chip device that creates the trapping potentials from current-carrying wires. We distinguish
between macroscopic current fluctuations generated by external
drivers (``technical noise'')
and Johnson noise, {\it i.e.,} microscopic fluctuations of thermal origin
in the metallic layers of the chip itself, which are typically at room temperature or higher. Johnson noise has a short correlation length~\cite{Carsten2003} and can
therefore cause direct loss of coherence over short length scales.
When the magnetic field fluctuates perpendicular to the
quantization axis
(along which the atomic spin is typically aligned), it may induce transitions between Zeeman sub-levels and
cause atoms to leave the trap, as discussed in more detail in Sec.~\ref{sec:magnoise}.
This loss mechanism may also cause dephasing, as we discuss in Sec.~\ref{sec:decoherence-by-loss}.
Although technical noise has a long correlation length, it may also contribute to dephasing through the latter mechanism.
However, as we discuss in Sec.~\ref{sec:magnoise}, technical noise may also lead to direct dephasing
if the corresponding current fluctuations
induce an asymmetric deformation in the trapping potential
as a result of summing
magnetic field vectors from nearby microwires and more distant sources.

\section{The two-mode model}
\label{sec:TMBH}

Our framework for the analysis of decoherence of a Bose gas in a double well is the two-site Bose-Hubbard model,
which is based on the assumption that all atoms occupy one of two spatial modes,
as described in the previous section. Before we analyze decoherence in the system we review
this model and its main predictions relevant to this work. The validity of the model for typical
scenarios presented in this paper is further discussed in Sec.~\ref{sec:validity}.

\subsection{Interaction and tunneling Hamiltonian}

Consider $N$ bosonic atoms in a double well such that two spatial modes may be occupied,
$\phi_L({\bf r})$ in the left well and $\phi_R({\bf r})$ in the right well. The dynamics is governed by the
two-site Bose-Hubbard  Hamiltonian~\cite{Sols1999,Leggett01,Gati07}
\begin{eqnarray} \hat{H} &=&
\frac{\epsilon}{2}(\hat{n}_L - \hat{n}_R)
-\frac{J}{2}(\hat{a}_L^{\dag}\hat{a}_R + \hat{a}_R^{\dag}\hat{a}_L) \nonumber \\
&& +
\frac{U}{2}
\sum_{j = L,R} \hat{n}_{j} (\hat{n}_{j} - 1)
\label{eq:Ham}
\end{eqnarray}
where $\epsilon$ is the energy imbalance (per particle)
between the two wells, $J$ is the tunneling matrix element and $U$ is the on-site interaction
energy per atom pair.
Here $\hat{a}_j$ ($j = L,R$)
are the bosonic annihilation operators of atoms in the two modes $\phi_j$ and $\hat{n}_j=\hat{a}_j^{\dag}\hat{a}_j$
are the corresponding number operators.

Since the Hamiltonian [Eq.~(\ref{eq:Ham})] conserves the total number of particles in the
two wells, we can write it in terms of
the pseudo-spin operators $\hat{S}_{1} = \frac{1}{2}(\hat{a}_L^{\dag}\hat{a}_R+\hat{a}_R^{\dag}\hat{a}_L)$,
$\hat{S}_{2} = -\frac{i}{2}(\hat{a}_L^{\dag}\hat{a}_R-\hat{a}_R^{\dag}\hat{a}_L)$ and $\hat{S}_{3} =
\frac{1}{2}(\hat{n}_L - \hat{n}_R)\equiv  \hat{n}$,
whereupon the Hamiltonian takes the form
\be \hat{H}=\epsilon\hat{S}_{3} - J\hat{S}_{1} + U\hat{S}_{3}^2. \label{eq:spinHam} \ee
Note that the total spin $\hat{S}_{1}^2 + \hat{S}_{2}^2 + \hat{S}_{3}^2
= s(s+1)$ is fixed by the total atom number $2s = N = \hat{n}_{L}
+ \hat{n}_{R}$, as it commutes with
the Hamiltonian in Eqs.~(\ref{eq:Ham})-(\ref{eq:spinHam}).
The Hilbert space of the two-mode model is spanned by the $\hat{S}_{3}$ eigenstates $\left | s, n\right\rangle$, corresponding to the number states basis $|n_L,n_R\rangle$ with
$n_{L,R} = N/2 \pm n=s\pm n$. It will prove advantageous to make use of dimensionless characteristic parameters which determine both stationary and dynamic properties; these are
\be u \equiv NU/J
\quad \hbox{and}\quad \varepsilon \equiv \epsilon/J
\,.
\ee

\subsection{Equivalence to the top and pendulum Hamiltonians}
The spin Hamiltonian in Eq.~(\ref{eq:spinHam}) is equivalent to a quantized top, whose spherical phase space is described by the conjugate non-canonical coordinates $({\hat\theta},{\hat\varphi})$ defined through
\begin{eqnarray}
\hat S_{3}  &=&
\frac{N}{2} 
\, \cos\hat{\theta},
	\label{eq:def-S3}\\
\hat S_{1}  &=&
\frac{N}{2} 
\, \sin\hat{\theta}\cos\hat{\varphi},
	\label{eq:def-S1}
\end{eqnarray}
in analogy to the definitions on the Bloch sphere.
Equation (\ref{eq:spinHam}) is thus transformed into the top Hamiltonian~\cite{Strzys08,Khripkov13},
\be
\hat{H}(\hat{\theta},\hat{\varphi}) \ = \
\frac{NJ}{2}\left[
\varepsilon \cos\hat{\theta}
- \sin\hat{\theta}\cos\hat{\varphi}
+ \frac{ u }{2} \cos^2\hat{\theta}
\right]\,.
\label{top}
\ee

If the populations in the two wells remain nearly equal during the dynamics, {\it i.e.,}  $\langle \hat S_3 \rangle \ll N/2$,
then the spherical phase space reduces to the ``equatorial
region'' around $\theta = \pi/2$,
where it may be described by the cylindrical coordinates
$\hat{n}$  and its canonically conjugate angle $\hat{\varphi}$, satisfying
$[\hat n, \, \hat\varphi] = i$.
The Hamiltonian [Eq.~(\ref{top})] is then approximated by the
Josephson pendulum Hamiltonian~\cite{Makhlin01}
\be
\hat{H}_J(\hat{n},\hat{\varphi}) \ = \
U \left(\hat{n}-n_\epsilon\right)^2 - \frac{1}{2}JN \cos\hat{\varphi}\,,
\label{JHam}
\ee
where $n_\epsilon = -\epsilon/2U$.
 
\subsection{Classical phase space and interaction regimes}
The classical states of the double-well system are SU(2) spin coherent states $|\theta,\varphi\rangle$~\cite{Tikhonenkov08}.
Dynamics restricted to  such states satisfies the GP mean-field approximation,
where all the atoms are in a single mode -- a superposition of $\phi_L$ and $\phi_R$ --
and ${\cal O}(1/N)$ fluctuations are neglected.
The operators in Eqs.~(\ref{eq:spinHam}), (\ref{top}), and (\ref{JHam}) are hence  replaced by corresponding real numbers. Thus, the GP approximation may be viewed as the classical limit of the quantum many-body Hamiltonian, with an effective Planck constant
$\hbar\rightarrow 2/N$.

\begin{figure}[t!]
      \centering
      \includegraphics[width=\columnwidth]{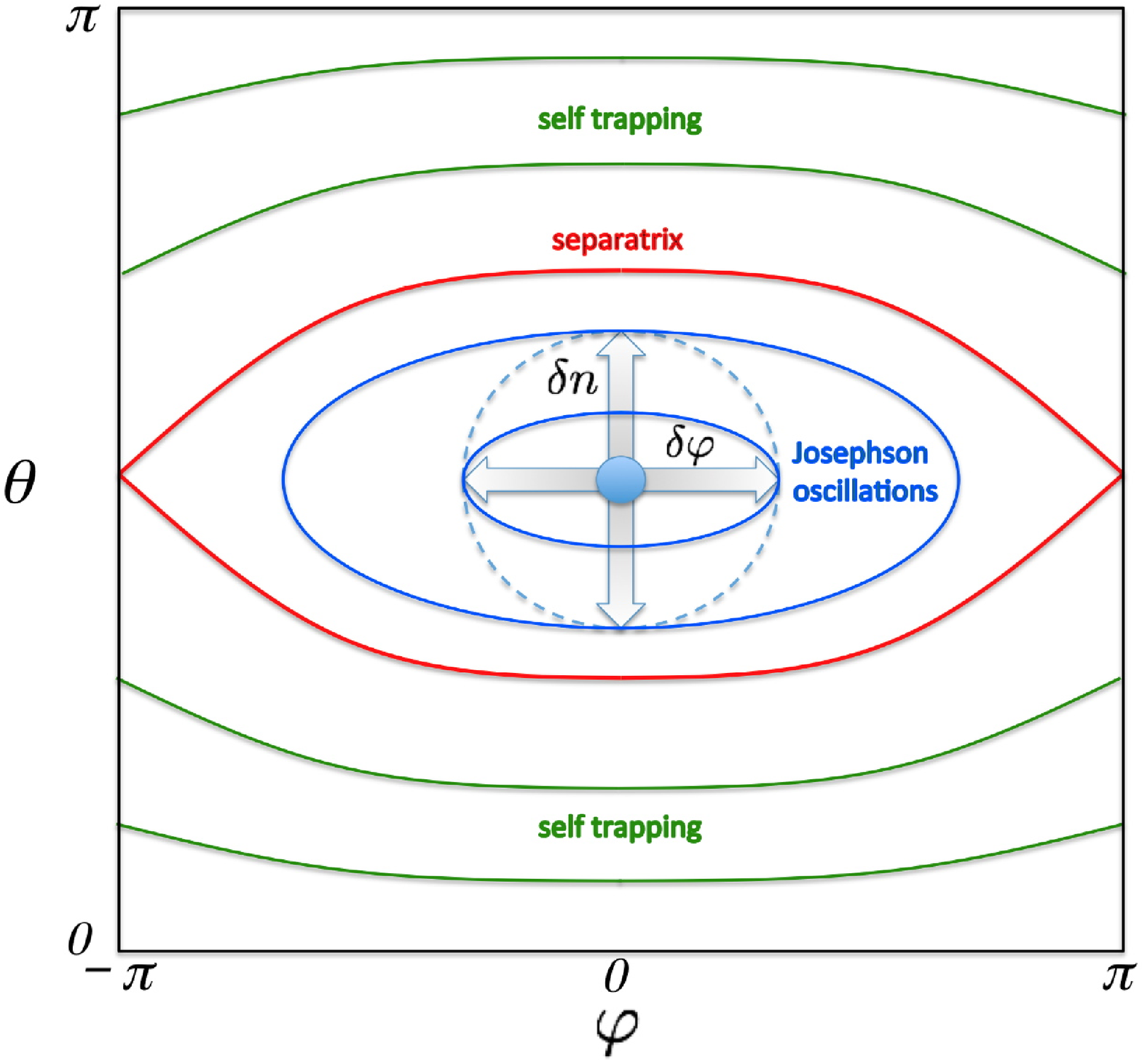} \\
\caption{Schematic illustration of iso-energy contours (solid lines) in the Josephson regime for a symmetric double-well potential ($\epsilon = 0$).
The spherical coordinates $\theta, \varphi$ map the Bloch sphere
of pseudospin operators $\hat{S}_\alpha$ ($\alpha = 1,2,3$)
[see Eqs.~(\ref{eq:def-S3})-(\ref{eq:def-S1})]: $\varphi$ is the
relative phase between the wells, and the left-right number
difference is proportional to $\cos\theta$.
The equilibrium distribution (circles or ellipses) are centered
on the $S_1$-axis.
The aspect ratio between the principal axes of the elliptical Josephson oscillation trajectories is $\xi^2$ [Eq.~(\ref{eq:def-squeezing})].
The dashed circle denotes an equal energy contour in the absence of interaction.
The small filled circle denotes the Gaussian-like phase-space distribution corresponding to the coherent
ground state (solid angle ${\cal O}(4\pi/N)$). The effect of stochastic rotations about the $S_{3}$ ($\delta \varphi$ kicks) and $S_{2}$ ($\delta n$ kicks)
axes is illustrated by arrows.  The combined action of kicking and nonlinear rotation is to spread the phase-space
distribution over the inner and outer ellipse, respectively.
}
\label{fig0}
\end{figure}

The qualitative features of the classical phase-space structure change drastically with the interaction strength~\cite{Chuchem10}.
If $|{\varepsilon}|<\varepsilon_c\equiv
\left(u^{2/3}-1\right)^{3/2}$, then for a strong enough interaction (${u>1}$)
two types of dynamics appear, which are described in the spherical $(\theta,\varphi)$ phase space by three
regions on the Bloch sphere, divided by a separatrix, as shown in Fig.~\ref{fig0}.
Motion within the region close to the equator (small population imbalance) is dominated by linear dynamics
and is characterized by Josephson oscillations of population between the wells. Conversely, motion between
 the separatrix and the poles is dominated by nonlinear dynamics and is characterized by
self-trapped phase oscillations with non-vanishing time-averaged population imbalance.

In what follows, we focus on the case of zero energy imbalance ($\epsilon=0$). Accordingly we distinguish between three regimes depending on the strength of the interaction~\cite{Gati07}:
\begin{eqnarray}
\mbox{Rabi regime:} &~& u<1~, \\
\mbox{Josephson regime:} &~& 1<u<N^2~,
	\label{ineq:Josephson}
	\\
\mbox{Fock regime:} &~& u>N^2~.
\end{eqnarray}
In the Rabi regime the separatrix disappears and the entire phase space is dominated by nearly linear population oscillations between the wells.
At the opposite extreme, in the Fock regime the nearly-linear region has an area less than $1/N$, and therefore it cannot accommodate quantum states.
Thus in the Fock regime, motion throughout the classical phase space corresponds to nonlinear phase oscillations with suppressed tunneling between the wells (self trapping).
In the absence of tunneling between the wells spatial coherence cannot be sustained even if external noise does not exist.
We will therefore focus on the dynamics within the  Josephson oscillations phase-space region, which exists in the Rabi and Josephson  interaction regimes.

\subsection{Semiclassical dynamics}
Beyond strictly classical evolution, much of the full quantum dynamics is captured by a {\em semiclassical} approach. This amounts to classical Liouville propagation of an ensemble of points corresponding to the initial Wigner distribution in phase space~\cite{Chuchem10}. Thus, while classical GP theory assumes the propagation of minimal Gaussians centered at $(\theta,\varphi)$ with fixed uncertainties (coherent
states), the semiclassical truncated-Wigner-like method permits the deformation of the phase-space distribution, thus accounting for its squeezing, folding, spreading, and dephasing. The only parts of the dynamics missed by such semiclassical evolution are true interference effects which only become significant for long time scales when different sections of the phase-space distribution overlap.
\subsection{Ground state and excitation basis}
As an alternative point of view that complements the semiclassical approach we also use a fully quantum
treatment of the two-mode Bose-Hubbbard model, which takes a simple analytic form when the
deviation from the minimum energy eigenstates of the Hamiltonian is small. For simplicity,
we will assume zero energy imbalance ($\epsilon=0$) and derive the structure of the ground state and lowest
energy excitations.

In the Fock regime ($u>N^2$) the ground state and low-lying excitations are number states. However, throughout the Rabi and Josephson regime ({\it i.e.,} for $u<N^2$) the ground state and low-lying excitations are nearly coherent and are characterized by a small phase uncertainty $\langle \hat{\varphi}^2\rangle\ll 1$.
We may therefore approximate the low-energy regime of the pendulum by a harmonic oscillator, {\it i.e.,} (ignoring the constant~$-1$) $-\cos\hat{\varphi}\to \hat{\varphi}^2/2$, converting the Josephson Hamiltonian [Eq.~(\ref{JHam})] to an oscillator in the canonical variables $\hat{n} \equiv \hat{S}_3$ and $\hat{\varphi}$,
\be \hat{H}_J\approx \frac{1}{2M}(\xi \hat{n})^2+\frac{M}{2}\omega_J^2(\hat{\varphi}/\xi)^2, \ee
where $\xi$ is the squeezing factor, given by
\be \xi=(1+u)^{1/4},
	\label{eq:def-squeezing}
\ee
$\omega_J$ is the Josephson frequency
\be \omega_J=\sqrt{J(J+NU)}=\xi^2 J, \ee
and $M\equiv N/2\omega_J$. The ground state of this Hamiltonian is Gaussian in $\xi \hat{n}$ and $\hat{\varphi}/\xi$,
and the excitation energies in this approximation are integer multiples of $\hbar\omega_J$.
More explicitly, we may define a bosonic operator
\be \hat{b}\equiv
\frac{ \sqrt{ N } }{ 2\xi } \hat{\varphi}
+ \frac{ i \xi }{ \sqrt{ N } } \hat{n}
~,
\label{bosop}\ee
such that
from $[\hat{\varphi}, \hat{n}]=i$ we obtain $[\hat{b},\hat{b}^{\dag}]=1$. The Josephson Hamiltonian can then be written
as
\be \hat{H}_J\approx \omega_J(\hat{b}^{\dag}\hat{b}+1/2),
\label{eq:normal-oscillator}
\ee
and the pseudo-spin operators of Eq.~(\ref{eq:spinHam}) are identified as
\begin{eqnarray}
\hat{S}_{1} &=& \frac{N}{2} \sin\hat{\theta}\cos\hat{\varphi}
\nonumber \\
&\approx & \frac{N}{2}-\frac{1}{4}\left[\xi^2(\hat{b}+\hat{b}^{\dag})^2-\xi^{-2}(\hat{b}-\hat{b}^{\dag})^2-2\right]
\label{eq:Sxbb} \\
\hat{S}_{2} &=& \frac{N}{2} \sin\hat{\theta}\sin\hat{\varphi}\approx \frac{1}{2}\sqrt{N}\xi (\hat{b}+\hat{b}^{\dag})
\label{eq:Sybb} \\
\hat{S}_{3} &=& \hat{n}\approx \frac{\sqrt{N}}{2i\xi}(\hat{b}-\hat{b}^{\dag})
~.
\label{eq:Szbb}
\end{eqnarray}
Here we have used a second-order expansion in $\hat{n}/N$ and
$\hat{\varphi}$ such that
$\sum_\alpha\hat{S}_{\alpha}^2 =\frac{N}{2}\left(\frac{N}{2}+1\right)$.

\subsection{Coherence}
\label{sec:TMBH_Coherence}

We are interested in the one-particle spatial coherence, namely the visibility of a fringe pattern
formed by averaging many events where the atoms are released from the double-well trap.
If we neglect experimental imperfections related to the release process or imaging, then the
coherence is the relative magnitude of the interference term of the momentum distribution
of the atoms in the trap~\cite{Pitaevskii2001}. If the two spatial modes are confined primarily to the left and right sites
of the potential, then the coherence is given by
\be g^{(1)}_{LR}=\frac{|\langle \hat{a}_L^{\dag}\hat{a}_R\rangle|}{\sqrt{n_L n_R}},
\label{eq:gLR} \ee
where $n_j\equiv \langle \hat{n}_j\rangle$ is the average number of atoms in the two modes.

For a symmetric double well ($\epsilon=0$) with equal populations of the two modes ($\langle \hat{S}_3 \rangle = 0$), the
definition of Eq.~(\ref{eq:gLR}) coincides with the normalized length $S/s
= 2 S / N$ of the
Bloch vector $\mathbf{S}\equiv (\langle{\hat S_{1}}\rangle, \langle{\hat S_{2}}\rangle, \langle{\hat S_{3}}\rangle)$.
Let us note that $s$ is the maximal allowed Bloch vector length ({\it e.g.,} after preparation of a coherent state), whereas $S$ is the actual Bloch vector length, {\it i.e.,} after time evolution under dephasing or if the prepared state was squeezed.
For the ground state preparations with $\langle\varphi\rangle=0$ considered below,
symmetry implies that the Bloch vector remains aligned along the $S_{1}$ axis, so that
$g^{(1)}_{LR}=S/s=\langle \hat{S}_{1}\rangle/s$
throughout the time evolution.

Semiclassically, for any Gaussian phase-space distribution,  the one-particle coherence is related to the variances $\Delta_a,\Delta_b$ of the distribution along its principal axes $a,b$
as~\cite{Khripkov12}
\be
\frac{S}{s}=\exp\left[-\frac{1}{2}\left(\Delta^2-2/N\right)\right],
\label{SSC}
\ee
where $\Delta^2=\Delta_a^2+\Delta_b^2$. A coherent state with isotropic $\Delta_a^2=\Delta_b^2=1/N$ thus has $S=s$, {\it i.e.,} $g^{(1)}_{LR}=1$.
Otherwise, for a squeezed Gaussian distribution with
$\Delta_a^2=\xi^2/N$, $\Delta_b^2=\xi^{-2}/N$, the coherence
drops to $g^{(1)}_{LR}=\exp[-\left[(\xi^2+\xi^{-2}-2)/2N\right]] < 1$.
For example, the ground state of the Josephson Hamiltonian is described by a Gaussian phase-space
distribution which is squeezed along the principal axes
with the squeezing factor
$\xi $ in Eq.~(\ref{eq:def-squeezing}), corresponding to a coherent state ($\xi=1$) only
in the non-interacting case ($u=1$).
More generally, time evolution such as the decoherence process that will be described in Sec.~\ref{sec:main} may lead to
a Gaussian distribution which has spread out by a factor $D$
so that $\Delta_a^2=D\xi^2/N$, $\Delta_b^2=D\xi^{-2}/N$ gives an even smaller value
$g^{(1)}_{LR}=\exp[-\left[D(\xi^2+\xi^{-2}-2)/2N\right]]$.

By using the approximation of Eq.~(\ref{eq:Sxbb}) for $\hat{S}_{1}$ in terms of the bosonic operator $\hat{b}$, one can see that
the coherence of the ground state is given by $g^{(1)}_{LR}\approx 1-(\xi^2+\xi^{-2}-2)/2N$, in agreement with the squeezed Gaussian semiclassical value. Thus, the ground state is coherent for the non-interacting case ($\xi=1$) and its coherence is reduced due to interactions. It is evident that in order to see significant reduction in the ground state coherence,  $\xi^2=\sqrt{u+1}$ should be comparable to $N$. Thus, the coherence of the ground state drops only in the transition from the Josephson to the Fock regimes ($u > N^2$)~\cite{Gati07}.  This crossover in the finite size system is the
simplest version of the superfluid to Mott insulator
transition~\cite{Bloch2008}.

The coherence drops further if excited states are populated.
This happens in thermal equilibrium at low temperatures, where the coherence factor is $g^{(1)}_{LR} \approx 1 -
[(\xi^2+\xi^{-2})(2n_T+1)-2]/2N$ with
$n_T\equiv \langle \hat{b}^{\dag}\hat{b}\rangle_T\approx [\exp(\hbar\omega_J/k_BT)-1]^{-1}$ being the thermal occupation of the excited levels. Reduction of
coherence also occurs due to external noise, as we shall see below.

\section{Dephasing and loss due to magnetic noise}
\label{sec:magnoise}

The interaction of the magnetic field fluctuations with an atom is given by the Zeeman Hamiltonian
\be \hat{V}_Z({\bf r},t) = -{\vec \mu}\cdot {\bf B}({\bf r},t), \ee
where ${\vec \mu}$ is the magnetic moment of the atom. If the magnetic field is not too strong then the
atom stays in a specific hyperfine state ($F$) and its magnetic moment is proportional to the angular momentum
${\bf F}$ through ${\vec \mu} = \mu_F {\bf F} = g_F\mu_B{\bf F}$,
$g_F$ being the Land\'e factor and $\mu_B$ the Bohr magneton.
For an ensemble of atoms with translational degrees of freedom, we adopt the language of second
quantization and
write the interaction Hamiltonian as
\be \hat{V}_Z( t ) = -\mu_F\sum_q \sum_{m,m'}\int d^3{r}
\hat{\cal F}_q^{mm'}({\bf r})B_q({\bf r},t),
\label{eq:Vmb} \ee
where the three operators
\be
\sum_{mm'} \hat{\cal F}_q^{mm'}\equiv
\sum_{mm'}
\hat{\psi}_m^{\dag}({\bf r})F_q^{mm'}\hat{\psi}_{m'}({\bf r})
	\label{eq:def-spin-density}
\ee
are the components of the magnetization density.
Here, $q$ labels the components of the magnetic field and the
angular momentum operator; the indices $m,m'$ label the Zeeman
states of the hyperfine level
($-F\leq m,m' \leq F$). The operators
$\hat{\psi}_{m}({\bf r})$ are field operators for atoms in the internal sublevel $|m\rangle$, satisfying
bosonic commutation relations $[\hat{\psi}_m({\bf r}),\hat{\psi}_{m'}^{\dag}({\bf r}')]
=\delta_{m,m'}\delta({\bf r}-{\bf r}')$.

In magnetic traps the Zeeman Hamiltonian [Eq.~(\ref{eq:Vmb})], with ${\bf B}$ as the trapping magnetic field,
determines the trapping potential [{\it e.g.,} Eq.~(\ref{eq:Vxyz})] for atoms in a Zeeman sublevel whose magnetic moment
is aligned parallel to the magnetic field. We define the quantization axis as the local direction parallel to the trapping
magnetic field ($\hat{p}$ direction).

We will now examine the effect of magnetic field fluctuations, such that
${\bf B}$ in Eq.~(\ref{eq:Vmb}) will represent changes in the magnetic field relative to an average trapping field.
These
are responsible either for fluctuations of the trapping potential (for parallel magnetic field components
$q=p$ or slowly varying fields in any direction) or for transitions between different Zeeman states (transverse components $q=\pm$
at the transition frequency for transitions with positive or negative angular momentum changes).

In this work we make two simplifying assumptions:
(a) only one spin component is trapped ($m=F$) and
once an atom is in another internal state it immediately escapes
from the trap;
(b) the atoms in the double-well
trap occupy one of two spatial states $\phi_L$ and $\phi_R$ corresponding to the two (left and right) wells.

For the interaction driven by the fluctuating
field $B_{p}$ parallel to the trapping magnetic field,
we only need the diagonal matrix element $F^{mm}_p = F$ ($m = F$).
Inserting the expansion of the field operator
$\hat\psi_F$ over the two (spatial rather than internal)
modes $\phi_{L,R}$,
we can therefore replace Eq.~(\ref{eq:def-spin-density}) with
\be \hat{\cal F}_{p}^{F,F} \to F\sum_{i,j=L,R}
\phi_i^*({\bf r})\phi_j({\bf r})\hat{a}_i^{\dag}\hat{a}_j. \label{eq:FFF} \ee
Note that in the presence of $B_p$ the Hamiltonian [Eq.~(\ref{eq:Vmb})] conserves the total number of trapped atoms. The
diagonal terms $i = j$ correspond to fluctuating energy shifts
that potentially dephase coherent superpositions of left and
right sites.

Magnetic fields perpendicular to the $\hat{p}$-axis drive spin flip
transitions
betwen the trapped level $m=F$
and the untrapped level $m=F-1$
(matrix element $F^{F-1,F}_- = \sqrt{F}$).
We can replace
Eq.~(\ref{eq:def-spin-density}) with
\be \hat{\cal F}_-^{F-1,F}\to \sqrt{F}
\sum_k\zeta_k^*({\bf r})\hat{c}_k^{\dag}
\sum_{j=L,R}\phi_j({\bf r})\hat{a}_j,
\ee
where $\zeta_k({\bf r})$ are the modes of the untrapped state
and $\hat{c}_k^{\dag}$ are the corresponding creation operators.
Note that slowly varying magnetic fields perpendicular to the average local quantization axis $\hat{p}$ would not
cause transitions to untrapped states because the atomic spin direction would adiabatically follow the local direction of the
magnetic field vector, such that the net effect of these slow fluctuations would be a change of the potential
in a manner similar to magnetic fluctuations parallel to $\hat{p}$.

We will now examine the master equation for the dynamics of the two processes: number-conserving processes
dominated by energy/phase fluctuations, and the loss processes.

\subsection{Dephasing}
\label{sec:3}

The number-conserving stochastic evolution due to a fluctuating
magnetic field aligned with the atomic magnetic moment
($\hat{p}$-axis)
is derived from the Hamiltonian [Eq.~(\ref{eq:Vmb})] keeping only
the component $B_{p}$ and the atomic operator given
in Eq.~(\ref{eq:FFF}).
The corresponding master equation for the density matrix
$\rho$ is of the form
$\dot{\rho}=i[\hat{H},\rho]+{\cal L}_N\rho$, where
${\cal L}_N$ is the number-conserving operator
\be
{\cal L}_N\rho=-\sum_{ijkl} \frac{\gamma_{N}^{ijkl}}{2}[\hat{a}_i^{\dag}\hat{a}_j\hat{a}_k^{\dag}\hat{a}_l\rho
+\rho\hat{a}_i^{\dag}\hat{a}_j\hat{a}_k^{\dag}\hat{a}_l
-2\hat{a}_i^{\dag}\hat{a}_j\rho\hat{a}_k^{\dag}\hat{a}_l]
\ee
and involves transition rates given by
\begin{eqnarray} \gamma_{N}^{ijkl} &=& \frac{\mu_F^2 F^2}{\hbar^2}\int d^3{r}\int d^3{r}'\,
\label{eq:dephasing-rate-1}
\\
&& \times \phi_i^*({\bf r})\phi_j({\bf r})
\phi_k^*({\bf r}')\phi_l({\bf r}')
{\cal B}_{pp}({\bf r},{\bf r}',\omega_{ijkl})
\,;
\nonumber
\\[6pt]
{\cal B}_{pp'}({\bf r},{\bf r}',\omega) &=&
\int d\tau e^{i\omega\tau} \langle B_{p}({\bf r},t)B_{p'}({\bf r}',t+\tau)\rangle
\,.
\end{eqnarray}
Here ${\cal B}_{pp}({\bf r},{\bf r}',\omega)$ is the two-point
correlation spectrum of the $p$-component of the magnetic field~\cite{Carsten1999}. The frequencies $\omega_{ijkl}\equiv
\omega_i+\omega_k-\omega_j-\omega_l$ are the transition frequencies, which may be taken to
be zero in our case since
we assume that the two modes $\phi_L$ and $\phi_R$ are degenerate. Here we also neglect energy shifts (magnetic
Casimir-Polder interaction) which are typically quite
small~\cite{Henkel2005,Haakh2009}.

If the spatial variation of the magnetic fields is small across
the trap volume of both sites, we may take
${\cal B}_{pp}({\bf r},{\bf r}',\omega_{ijkl})$
in Eq.~(\ref{eq:dephasing-rate-1}) outside the double integral.
The orthogonality relations among the
modes $\phi_j$ imply
that only the terms with $i=j$ and $k=l$ survive.
We then obtain $\gamma^{ijkl}_{N} = \delta_{ij}\delta_{kl}
\gamma_{N}$ with
\begin{equation}
\gamma_{N} =
(\mu_FF/\hbar)^2
{\cal B}_{pp}({\bf r}_t,{\bf r}_t,0),
\label{eq:dephasing-rate}
\end{equation}
where ${\bf r}_t$ is a typical position in the trapping region.
The dissipative term in the
master equation becomes
\be {\cal L}_N\rho=-\frac{\gamma_{N}}{2}\sum_{i,j=L,R}[n_in_j\rho+\rho n_i n_j-2n_i\rho n_j].
\label{eq:Lrho-const}
\ee
If the total number of atoms $n_L+n_R=N$ is fixed, this term vanishes completely.

If ${\cal B}_{pp}({\bf r},{\bf r}',0)$ varies over the trap region, then $\gamma_N^{ijkl}$ may be non-zero for any set of indices.
However, if the modes $\phi_L({\bf r})$ and $\phi_R({\bf r})$ are well separated, having only a small overlap
in the region of the barrier, then the terms with $i=j$ and $k=l$ are much larger.
These terms may be written in the form
\begin{eqnarray}
\gamma_{N}^{iijj} &=& \frac{\mu_F^2F^2}{\hbar^2}{\cal B}_{pp}({\bf r}_t,{\bf r}_t,0)\alpha_{ij}
\nonumber
\\[6pt]  
\alpha_{ij} &=& \int d^3{r}\int d^3{r}' {\cal A}({\bf r},{\bf r}')|\phi_i({\bf r})|^2|\phi_j({\bf r}')|^2\,,
	\label{eq:alpha-ij}
\end{eqnarray}
where ${\bf r}_t$ represents a typical location in which ${\cal B}_{pp}$ may be maximal, and
the dimensionless function ${\cal A}({\bf r},{\bf r}')$ represents the spatial shape of ${\cal B}_{pp}$ relative to its
value at ${\bf r}_t$.
If all elements of the (symmetric) matrix $\alpha_{ij}$ were equal,
they would give no contribution to the master equation,
similar to the argument in Eq.~(\ref{eq:Lrho-const}).
By subtracting $\alpha_{LR}$ from all elements,
we obtain the following dissipative terms in the master
equation
\begin{eqnarray} {\cal L}_N\rho  &=& -\frac{\gamma_N}{2}
\sum_{j = L, R} (\alpha_{jj} -\alpha_{LR})[n_j^2\rho+\rho n_j^2-2n_j\rho n_j] \nonumber \\[6pt]
&=& -\gamma_{p}(\hat{S}_{3}^2\rho+\rho\hat{S}_{3}^2-2\hat{S}_{3}\rho\hat{S}_{3}),
\label{eq:masterS3}
\end{eqnarray}
where
\be \gamma_p=\frac{\gamma_N}{2}(\alpha_{LL}+\alpha_{RR}-2\alpha_{LR})
\label{eq:gamma_p} \ee
and we have used $\hat{n}_{L,R}=N/2\pm \hat{S}_3$ with
the total number $N$ being conserved.
As expected, magnetic fluctuations along the quantization ($\hat{p}$)
axis apply phase noise between the two wells. This can be
represented by random rotations about the $S_3$-axis of the Bloch
sphere, and the dissipative term [Eq.~(\ref{eq:masterS3})] generates
the corresponding phase diffusion in the density operator.

Let us now consider two typical types of noise. The first is Johnson noise from thermal current fluctuations
near a conducting surface.
Let us focus for simplicity on a double-well potential whose sites
are equally close to the surface. The correlation function
${\cal B}_{pp}({\bf r},{\bf r}')$ then depends only on the
distance $|{\bf r} - {\bf r}'|$ and decays to zero on a length
scale $\lambda_c$ comparable to the height of the trap above the surface~\cite{Carsten2003}. Consider for example the simple
model ${\cal B}_{pp}({\bf r},{\bf r}')
\propto \exp(-|{\bf r}-{\bf r}'|/\lambda_c)$.
When the correlation length $\lambda_c$ is smaller than the
distance $d$ between the two sites, the off-diagonal terms
in the matrix $\alpha_{ij}$ [Eq.~(\ref{eq:alpha-ij})]
decay like $\alpha_{LR}\sim e^{-d/\lambda_c}$, while the diagonal
terms scale like
$\alpha_{jj}\sim V_c/V_{\phi_j}$, which is the fraction
of the mode volume ($V_{\phi_j}$) lying within a radius
$\lambda_c$ around the trap center. For very short correlation
lengths, $V_c\sim \lambda_c^3$.
It follows that when the correlation length is smaller than the
distance $d$, the off-diagonal terms $\alpha_{LR}$ become small
and a dephasing process takes place. This conclusion also
applies when the correlation function
${\cal B}_{pp}({\bf r},{\bf r}')$ is not exponential,
but Lorentzian, as computed in Ref.~\cite{Carsten2003}.

The second example is noise with a long correlation length that changes the effective magnetic potential, as
happens typically with so-called ``technical noise". Assume that one (or more) of the parameters
in the potential of Eq.~(\ref{eq:Vxyz}), such as $\omega_x$, $V_0$ or $d$ is fluctuating, or that another fluctuating
term such as $\delta V(t)=f(t)x$ is added to the potential.
If any of these variables (denoted by a generic name $v$)
fluctuates as $\delta v(t)$, where $\langle \delta v\rangle=0$ and $\int d\tau \langle \delta v(t)\delta v(t+\tau)\rangle=\eta>0$,
then the magnetic field correlation function has the form
${\cal B}_{pp}({\bf r},{\bf r}',0)\propto (\partial V({\bf r})/\partial v)(\partial V({\bf r}')/\partial v)\eta$
 so that
${\cal A}({\bf r},{\bf r}')$ in Eq.~(\ref{eq:alpha-ij}) can be factorized into a product of a function of ${\bf r}$ and the same function at ${\bf r}'$.
This implies that $\alpha_{ij}$ can be also factorized as $\alpha_{ij}=\beta_i \beta_j$,
where $\beta_j\propto \int d^3 r\ (\partial V({\bf r})/\partial v)|\phi_j({\bf r})|^2$.
We then get a dephasing rate $\gamma_p\propto (\beta_L-\beta_R)^2$.
It follows that for noise of technical origin, dephasing is expected
whenever the potential changes due to the magnetic fluctuations are asymmetric. For example,
if the magnetic field fluctuations create a linear slope,
$\delta V({\bf r}) = f(t) x$, then
with $\langle f(t)f(t')\rangle=\eta\delta(t-t')$ we obtain $\gamma_p=\frac{1}{2}(d/\hbar)^2\eta$, where $d$ is the
distance between the centers of the two wells. This is indeed what we expect for white phase noise $\delta\varphi(t)\sim f(t)d/\hbar$
between the two sites.

If the two wells are not fully separated, then we should expect other terms in the master equation, which
will typically reduce the dephasing rate. However, for magnetic
fields with short correlation lengths, we should also
expect the appearance of cross terms $\hat{a}_L^{\dag}\hat{a}_R$
in the master equation. These are driven by spatial gradients
of the magnetic field (see Refs.~\cite{Carsten2001,Henkel1999a}
for a more detailed discussion). 
On the Bloch sphere, cross terms correspond to random rotations around the $S_1$-axis (tunneling rate fluctuations) and around the $S_2$-axis (number noise).
For completeness, we include such terms in the following discussion.
Although they are expected to be small,
they may be amplified by atomic collisions,
as we shall see below.

\subsection{Loss}

The magnetic fields transverse to the quantization axis
lead to a loss interaction
\be \hat{\cal V}_{\rm loss} =
\sum_k \hat{c}_k^{\dag}
[g_{kL}(t)\hat{a}_L+g_{kR}(t)\hat{a}_R]
+ {\rm h.c.}, \ee
where
\be g_{kj}(t)=-\mu_F\sqrt{F}\int d^3{r} B_-({\bf r},t)\zeta_k^*({\bf r})\phi_j({\bf r})
\,, \ee
and $B_-$ is a complex-valued circular component
that lowers the angular momentum by one unit
[$B_- = (B_x + {\rm i} B_y)/\sqrt{2}$
if the quantization axis is along $z$].
The relevant loss term in the master equation for the atoms
in the two wells is then given by
\be {\cal L}_{\rm loss}\rho=-\sum_{i,j=L,R}\frac{\gamma^{ij}_{\rm loss}}{2}[\hat{a}_i^{\dag}\hat{a}_j\rho
+\rho\hat{a}_i^{\dag}\hat{a}_j-2\hat{a}_i\rho\hat{a}_j^{\dag}], \label{eq:mastera} \ee
where the loss rates are given by
\begin{eqnarray} \gamma^{ij}_{\rm loss} &=& \frac{1}{\hbar^2}\sum_k \int d\tau e^{i\omega_k\tau}\langle g_{ki}(t) g^*_{kj}(t+\tau)\rangle
\nonumber \\[6pt]
&=&\frac{\mu_F^2 F}{\hbar^2}
\sum_k \int d^3{r}\int d^3{r}'   \\
&&\times\ \zeta_k({\bf r})\phi_i^*({\bf r})
\phi_j({\bf r}')\zeta_k^*({\bf r}')
{\cal B}_{-+}({\bf r},{\bf r}',\omega_k)
\nonumber
\\[6pt]
{\cal B}_{-+}({\bf r},{\bf r}',\omega)
&=& \int d\tau e^{i\omega\tau} \langle B_-({\bf r},t)B_+({\bf r}',t+\tau)\rangle\,,
\end{eqnarray}
where
${\cal B}_{-+}({\bf r},{\bf r}',\omega)$
is the spectral density of transverse magnetic field fluctuations,
$B_+ = B_-^*$,
and $\omega_k$ contains the Zeeman and kinetic energies of the
non-trapped level.

If we assume that the magnetic spectrum is flat over the range
of energies $\omega_k$, then we can use the completeness relation
of the non-trapped states
$\sum_k \zeta^*_k({\bf r})\zeta_k({\bf r}')=
\delta({\bf r}-{\bf r}')$ to obtain
\be \gamma^{ij}_{\rm loss} = \frac{\mu_F^2 F}{\hbar^2}
\int d^3{r}\phi_i^*({\bf r})\phi_j({\bf r})
{\cal B}_{-+}({\bf r},{\bf r},\omega_Z),
\ee
where $\omega_Z$ is the average transition energy to the untrapped
Zeeman level.
If the magnetic noise spectrum depends weakly on the position
${\bf r}$ over the trap region, the orthogonality relations
between the modes
$\phi_L$ and $\phi_R$ imply that
the off-diagonal loss rates vanish and we are left with
\be \gamma^{ij}_{\rm loss} = \delta_{ij}\frac{\mu_F^2 F}{\hbar^2}
{\cal B}_{-+}({\bf r}_t,{\bf r}_t,\omega_Z),
\label{eq:loss-rate}
\ee
where ${\bf r}_t$ represents the region occupied by the atoms.

Compared to the dephasing rate $\gamma_{p}$
[Eq.~(\ref{eq:gamma_p})] of the previous subsection,
loss is harder to suppress: it occurs whenever the spectrum of the magnetic fluctuations has significant
components at the transition frequency $\omega_Z$ and does not depend on the correlation length of the field.
However, for Johnson noise at trap-surface distances of the same order as the distance between the wells,
the two rates happen to be similar
since the magnetic noise spectrum
is typically quite flat in frequency
and not strongly anisotropic~\cite{Carsten1999}.
Let us also note that several suggestions exist on how to suppress the overall Johnson noise ({\it e.g.,} Ref.~\cite{EurPhysJD35-87}), and in addition, how to suppress specific components of the field, such as those contributing to $\gamma_p$ ~\cite{David2008}. For experimental measurements of loss rates, see Sec.~\ref{sec:lifetime}.

\section{Combined dynamics}
\label{sec:main}

In order to investigate the combined dynamics of tunneling and interactions in the presence of noise we make
some analytical estimations of the expected decoherence rate and also solve the master equation numerically
for specific configurations. We assume that the atomic gas
is initially in the lowest energy
eigenstate for a specific number of atoms $N$. Usually $N=50$
in our numerical simulations, and both sites are symmetric
with $N/2$ occupation on average. We then
turn on the noise source and follow the coherence of the Josephson junction as a function of time.

\subsection{Number conserving noise (no loss)}
\label{sec:main_A}

For visualizing the combined effects of the nonlinear Hamiltonian
dynamics of Sec.~\ref{sec:TMBH} and the magnetic noise,
a semiclassical picture involving the distribution over the
spherical phase space (Bloch sphere) is illuminating. For
example, the master equation generated by the dissipative
term of Eq.~(\ref{eq:masterS3}) can be represented by adding
a term $f_3(t){\hat S}_3$ to the Hamiltonian where $f(t)$ is
an erratic driving amplitude (a random process with zero average
and short correlation time~\cite{Khripkov12}).
This may be viewed as a realization of a Markovian stochastic process such that upon averaging,
\be
\langle f_3(t)f_3(t')\rangle = 2\gamma_3\delta(t-t')\,.
\ee
On the Bloch
sphere, such a driving term corresponds to a random sequence
of rotations around the ``north-south'' (or $S_3$-) axis.
The phase-space distribution in Fig.~\ref{fig0} then diffuses
in the $\varphi$-direction so that the relative phase between
the left and right wells gets randomized (phase noise).
The parameter $\gamma_3$
is the corresponding phase diffusion rate.
This is consistent with the fact that the additional Hamiltonian
corresponds to a random bias of the double-well potential,
\begin{equation}
f_3(t){\hat S}_3 = \frac{ f_3(t) }{ 2 } (\hat{n}_L - \hat{n}_R)\,.
\label{eq:interp-S3-drive}
\end{equation}
Similarly, we also consider random rotations around
other axes of the Bloch sphere, for example
\begin{equation}
f_2(t) \hat{S}_2 = - i \frac{ f_2(t) }{ 2 }
(\hat{a}_L^\dag \hat{a}_R - \hat{a}_R^\dag \hat{a}_L)\,.
\label{eq:S2-drive}
\end{equation}
This corresponds to inter-site hopping with a
random amplitude and is generated by fluctuating
inhomogeneous magnetic fields (Sec.~\ref{sec:magnoise}).
On the Bloch sphere of Fig.~\ref{fig0},
we then have, near $\varphi=0$, random kicks that spread the phase-space
distribution along the ``north-south'' (or $\theta$) direction.
They change the population difference which is proportional (for $|\theta-\pi/2|\ll 1$)
to $\cos\theta \approx \pi/2 - \theta$, and we denote $\gamma_2$ as
the corresponding diffusion rate (number noise).
Note that the effect of rotations around the $\hat{S}_1$-axis [involving a $+$ sign on the right-hand side of Eq.~(\ref{eq:S2-drive})] on phase-space points near $\varphi=0,\theta=\pi/2$ is more intricate. It amounts to random fluctuations of the tunneling rate $J$ in the Hamiltonian [Eq.~(\ref{eq:Ham})] and will be discussed briefly at the end of this subsection.  
%

The actual dynamics must
also consider the other parts of the Bose-Hubbard
Hamiltonian [Eq.~(\ref{eq:Ham})].
Consider first the interaction-free case $U = 0$ with the squeezing
parameter $\xi=1$. Classical trajectories in the vicinity of the ground state are perfect circles
on the Bloch sphere
centered on the $S_{1}$-axis.
When the Rabi-Josephson rotation is faster than the diffusion/kicking rate, both types of noise above produce isotropic 2D diffusive spreading of the phase-space distribution, with a diffusion rate $\gamma_\alpha$ ($\alpha = 2,3$). Thus, at large times $t$, one obtains a circular Gaussian phase-space distribution whose radial variance grows linearly as $\Delta^2=2\gamma_\alpha t$. Substitution into Eq.~(\ref{SSC}) gives the expected exponential decay $S/s\propto \exp{[-\gamma_\alpha t]}$.
This regime flattens out when the state has spread over the
entire Bloch sphere, $\Delta \sim \pi$.

Now, when interactions are present, it is evident that the two types of noise will give different results.
The Josephson trajectories are squeezed, with a $1/\xi^2$ ratio between their principal axes~(Fig.~\ref{fig0}).
The combined action of phase noise ($\alpha = 3$) and Josephson rotation gives an elliptical distribution whose variances are $\Delta_2^2 = 2\gamma_{3} t$ and $\Delta_{3}^2=2\gamma_{3} t/\xi^4$ (inner blue ellipse in Fig.~\ref{fig0}). In contrast, for number noise ($\alpha = 2$), the distribution at time $t$ has variances $\Delta_2^2 = 2\xi^4 \gamma_{2} t$ and $\Delta_{3}^2=2\gamma_{2} t$ (outer blue ellipse in Fig.~\ref{fig0}).
Therefore, using Eq.~(\ref{SSC}) the one-particle coherence decays
as
\be
S/s \propto \exp {\left(-\Gamma_{3,2} t\right)}\,,
\label{Sdecay}
\ee
with effective decay rates
\be
\Gamma_{3,2} = \frac{\gamma_{3,2}}{2}\left(1+\xi^{\mp 4}\right).
\label{effrate}
\ee
Thus, for phase (number) noise the decoherence rate is suppressed (enhanced) in the presence of interactions with respect to the interaction-free case, respectively.
The best suppression factor one can expect in the former case is a factor of two,
{\it i.e.,} $\Gamma_{3} = \gamma_{3}/2$.
This limiting factor of~$2$ in the suppression can be understood in the following way: in Fig.~\ref{fig0} the interactions can restrict the vertical spreading but not the horizontal spreading. This evolution then leads to a distribution with a relatively small $\Delta n$ but a large $\Delta \varphi$. In other words, observing the dashed circle and inner ellipse in Fig.~\ref{fig0}, no matter how much interactions shrink the ellipse, its $\Delta\varphi$ width will be the same as it was without interactions. The coherence of a distribution tightly squeezed along the equatorial line with this $\Delta\varphi$ is precisely half that of the dashed circle.
It should be noted that our model does not take into account
processes that relax the system towards its ground state and
actually create phase coherence.

In the derivation of Eq.~(\ref{effrate}), it was assumed that the Josephson oscillations spread the semiclassical ensemble throughout the pertinent classical orbit. 
However, it may happen that diffusion due to the 
noise leads to an ensemble with an oscillating width. This is particularly noticeable at short times and if the noise is switched on faster than the Josephson period.
The combination of anisotropic diffusion and motion along squeezed elliptical
classical trajectories then leads to a ``breathing'' ensemble and a coherence that oscillates
with half the Josephson period about the decaying coherence value of Eq.~(\ref{Sdecay}), as we show below.

For a more quantitative analysis, we use a master equation with a dissipative term of the structure of Eq.~(\ref{eq:masterS3}) to derive rate
equations for the relevant bilinear observables of the Josephson junction: the occupation $b^\dag b$ and the ``anomalous occupation'' $bb$.
Let us begin with number noise where the $\hat{S}_{3}$
operator is replaced by $\hat{S}_{2}$, according to the
random rotation picture of the preceding paragraph.
We obtain the dynamical equations
\begin{eqnarray} \frac{d}{dt}\langle \hat{b}^{\dag}\hat{b}\rangle &=&  \gamma_{2}\frac{N\xi^2}{4} \\
\frac{d}{dt} \langle\hat{b}^2\rangle &=& -2i\omega_J\langle\hat{b}^2\rangle-\gamma_{2}\frac{N\xi^2}{4}~, \end{eqnarray}
where $\gamma_2$ is the diffusion rate related to number noise.
This has the simple solution,
\begin{eqnarray} \langle \hat{b}^{\dag}\hat{b}\rangle_t &=& \langle \hat{b}^{\dag}\hat{b}\rangle_0
+\gamma_{2}\frac{N}{4}\xi^2 t \\
\langle \hat{b}^2\rangle_t &=& \langle \hat{b}^2\rangle_0 e^{-2i\omega_J t}
+i\frac{\gamma_{2} N\xi^2}{8\omega_J}(1-e^{-2i\omega_J t})\,. \end{eqnarray}
Substitution into Eq.~(\ref{eq:Sxbb}) gives the coherence evolution for an initial state where $\langle \hat{b}^2\rangle_0 = 0$
(this applies to the ground state and in thermal equilibrium),
\be \langle \hat{S}_{1}\rangle_t= \langle \hat{S}_{1}\rangle_0
-\frac{\gamma_{2} \xi^2}{4}\left[(\xi^2+\xi^{-2})t-\frac{\sin 2\omega_J t}{2\omega_J} (\xi^2-\xi^{-2})\right],\ee
implying that
\be \Gamma_{2}(t)\equiv -\frac{\partial}{\partial t}\langle \hat{S}_{1}\rangle_t=\frac{\gamma_{2}\xi^2}{2}\left[\xi^{-2}
\cos^2\omega_J t+\xi^2 \sin^2\omega_J t\right].
\label{Gammaydecay}
\ee
Thus, the decay rate oscillates between $\gamma_{2}$ and $\gamma_{2}\xi^4$ with the average rate
given by $\gamma_{2} (1+\xi^4)/2$, as in Eq.~(\ref{effrate}).

Similarly we obtain for phase noise (random kicks around
the $S_{3}$-axis),
\be \Gamma_{3}(t)=\frac{\gamma_{3}\xi^{-2}}{2}\left[\xi^{2}
\cos^2\omega_J t+\xi^{-2} \sin^2\omega_J t\right], \label{Gammazdecay}\ee
implying oscillations of the decay rate between $\gamma_{3}$ and $\gamma_{3}\xi^{-4}$ with an average rate of $\gamma_{3}(1+\xi^{-4})/2$.

\begin{figure}[t!]
      \centering
      \includegraphics[width=\columnwidth]{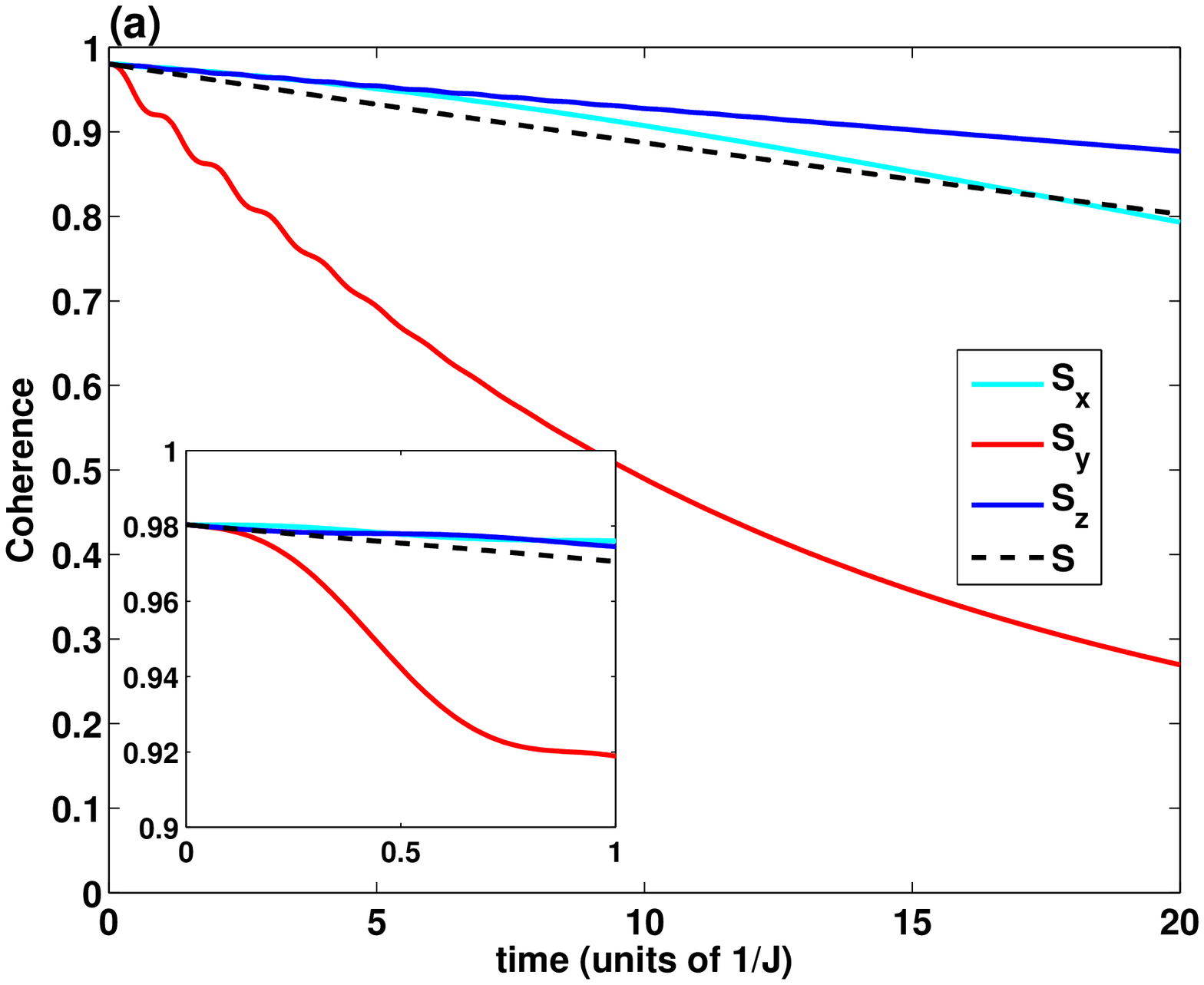} \\
      \includegraphics[width=\columnwidth]{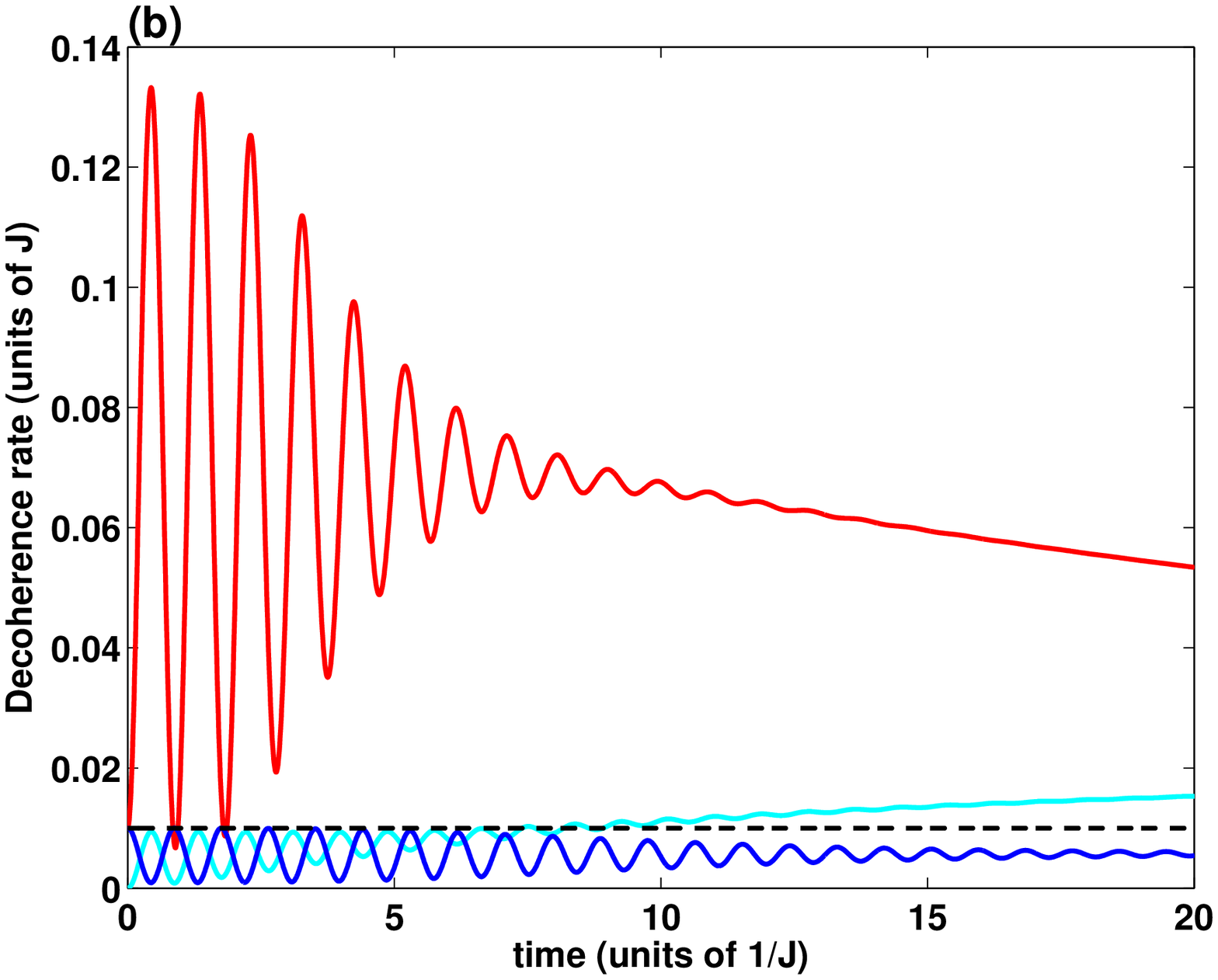}
 \caption{Decoherence process due to number-conserving noise in a double well. Time is scaled to the inverse of the tunneling rate $J$,
and the on-site interaction energy in the Hamiltonian [Eqs.~(\ref{eq:Ham}) and~(\ref{eq:spinHam})] is $U=0.25\,J$ with $N=50$ particles. The Josephson frequency is
$\omega_J=3.67\,J$, corresponding to a squeezing factor $\xi=1.92$. 
(a) coherence as a function of time for random $S_{1}$, $S_{2}$ and $S_{3}$ rotations
with corresponding stochastic rates $\gamma_{\alpha}=\gamma=0.01\,J$.
For reference, we also plot the single particle
decoherence where $g^{(1)}_{LR}\propto e^{-\gamma t}$ (dashed line).
The inset shows the short time behavior. (b) instantaneous coherence decay rate for the 3 cases,
showing oscillations with approximately the Josephson frequency. The dashed line again shows the single-particle case for reference.
}
\label{fig1}
\end{figure}

The predictions of Eqs.~(\ref{Sdecay}), (\ref{Gammaydecay}), and~(\ref{Gammazdecay}) are confirmed by exact numerical solutions of the master equation with the dissipative term having the form of Eq.~(\ref{eq:masterS3}).
As we showed in Sec.~\ref{sec:3}, magnetic noise in a double-well trap leads mainly to phase noise represented
by an erratic field proportional to $\hat{S}_{3}$. However, in the spirit of the general discussion above and for
completeness of the discussion, we have also
calculated the evolution of the coherence for $\hat{S}_{2}$ and $\hat{S}_{1}$
as driving forces: random rotations around the
$\hat{S}_{2}$-axis that generate number
noise, and around the  $\hat{S}_{1}$-axis that generate tunneling-rate fluctuations.
In Fig.~\ref{fig1} we present the evolution of coherence for given parameters (see caption) and different rotation axes.

The decay rates $\Gamma_{\alpha}(t)\equiv -\partial/\partial t[\log \langle \hat{S}_{1}\rangle_t]$
for the noise rotating around $\hat{S}_{\alpha}$ are presented in Fig.~\ref{fig1}b ($\gamma_{\alpha}=\gamma$ for $\alpha = 1, 2, 3$).
The decoherence rate
for rotations around $\hat{S}_{1}$ (cyan curve) oscillates between $\Gamma_{1}=0$ and $\Gamma_{1}=\gamma$ and then starts
to increase gradually. 
The decoherence rate $\Gamma_{3}$ (phase noise, blue curve) oscillates between $\gamma$ and $\gamma/\xi^4$
and then stabilizes with the average value $\Gamma_{3}\approx \gamma(1+\xi^{-4})/2$ of Eq.~(\ref{effrate}).
The fastest decay of coherence is found for number noise (red curve, $\alpha = 2$) because random rotations around this axis significantly displace the squeezed ensemble from its equilibrium position.
The rate
first oscillates between $\gamma$ and $\gamma\xi^4$ and then stabilizes with the average value
$\Gamma_{2}=\gamma(1+\xi^4)/2$.
It gradually decreases to smaller values when the coherence is already small. 

The decay of the decoherence rate oscillations may be interpreted within the semiclassical approach as 
following from the dispersion of Josephson periods for classical trajectories with different perimeters. 
In the linearized excitation approach, this corresponds to 
unequally spaced energy levels due to the deviation from the harmonic approximation.
Note that similar oscillations have been observed in Ref.~\cite{Henkel2004}
for an oscillator and a certain scenario of spatial decoherence.

The decay of the coherence in the case of tunneling rate fluctuations
(random rotations around $\hat{S}_{1}$), which leave the system
invariant in the absence of interactions (circular distribution
in Fig.~\ref{fig0}), can also be explained by the squeezing.
The rotation of an elliptical ground state changes it
and excites higher energy
states (or classical trajectories). After some time these excitations become similar to number excitations and
the decay rate increases.


%

\subsection{Decoherence induced by loss}
\label{sec:decoherence-by-loss}

Loss of atoms from traps due to noise-induced transitions to untrapped internal atomic states is a very common
process in atom chip traps. Here we examine the possibility that a loss process leads not only to the reduction of the
total number of atoms in the trap, but also to decoherence of the remaining BEC. Let us begin by noting that while a loss process
does not usually heat up a BEC by transitions into higher-energy trap levels, it does increase the uncertainty in the number
of remaining atoms. If we imagine two BECs in two
separate traps which initially have a fixed number of atoms, then independent loss from both traps will lead to
uncertainty in the relative numbers in the two traps.
Due to interactions, this makes the chemical potential uncertain
and leads to dephasing if the phase between the two BECs is initially well defined. Does this process of independent loss from the two
traps, which is described by Eq.~(\ref{eq:mastera}), indeed lead to decoherence as
we have shown above for number noise? 

Consider a trapped Bose gas of $N$ atoms in a double well. Atom loss from the left (right) well is described by the
application of the annihilation operator $\hat{a}_L$ ($\hat{a}_R$) to the system state. In the resulting state with $N-1$
atoms, the expectation value of a pseudo-spin operator is
\be \langle \hat{S}_\alpha\rangle_{N-1}
=
\frac{\langle \hat{a}_{i}^{\dag}\hat{S}_\alpha\hat{a}_{i}\rangle_N}{\langle \hat{n}_{i}\rangle_N}\
\ee
for $i = L,R$.
Specifically we have
\be \langle \hat{S}_{1}\rangle_{N-1}\approx \frac{N-1}{N}\langle \hat{S}_{1}\rangle_N\pm
\frac{\langle\hat{S}_{3}\hat{S}_{1}\rangle+\langle\hat{S}_{1}\hat{S}_{3}\rangle}{N}
\ee
\be \langle \hat{S}_{3}\rangle_{N-1}
\approx \langle \hat{S}_{3}\rangle_N\mp \frac{1}{2}\left(1-\frac{4\langle \hat{n}^2\rangle_N}{N}\right), \ee
where the $+$ and $-$ refer to $L$ and $R$, respectively.
We have dropped terms of order $\langle \hat n\rangle/N$, assuming nearly equilibrated populations in the wells.
If the initial state is the ground state of non-interacting atoms,
it is also a eigenstate (coherent state) of
$\hat{S}_{1}$, and we have $\langle \hat{n}^2\rangle_N=N/4$.
The application of $\hat{a}_L$ or $\hat{a}_R$ does not change the coherence
nor the number difference: the state remains a coherent state of $N-1$ atoms.
However, if the $N$-atom state is the squeezed ground state of an interacting system,
then $\langle \hat{n}^2\rangle_N=N/4\xi^2$ and each atom lost from either well changes the
imbalance between the two wells by $\pm (1-1/\xi^2)$. It then follows that the loss process from an interacting
system is equivalent to random kicks in the number direction,
similar to the number noise considered above.
Then the decoherence rate would be
\be \Gamma_{\rm dec}\sim \gamma_{\rm loss}\frac{1-1/\xi^2}{2N}[c(t)+s(t)\xi^4]\,,
\label{eq:gamma_loss} \ee
where $c(t)$ and $s(t)$ are functions that we would expect to behave like $\cos^2\omega_J t$ and $\sin^2\omega_Jt$
for short times and then stabilize with their average value $1/2$.

\begin{figure}[t!]
      \centering
      \includegraphics[width=\columnwidth]{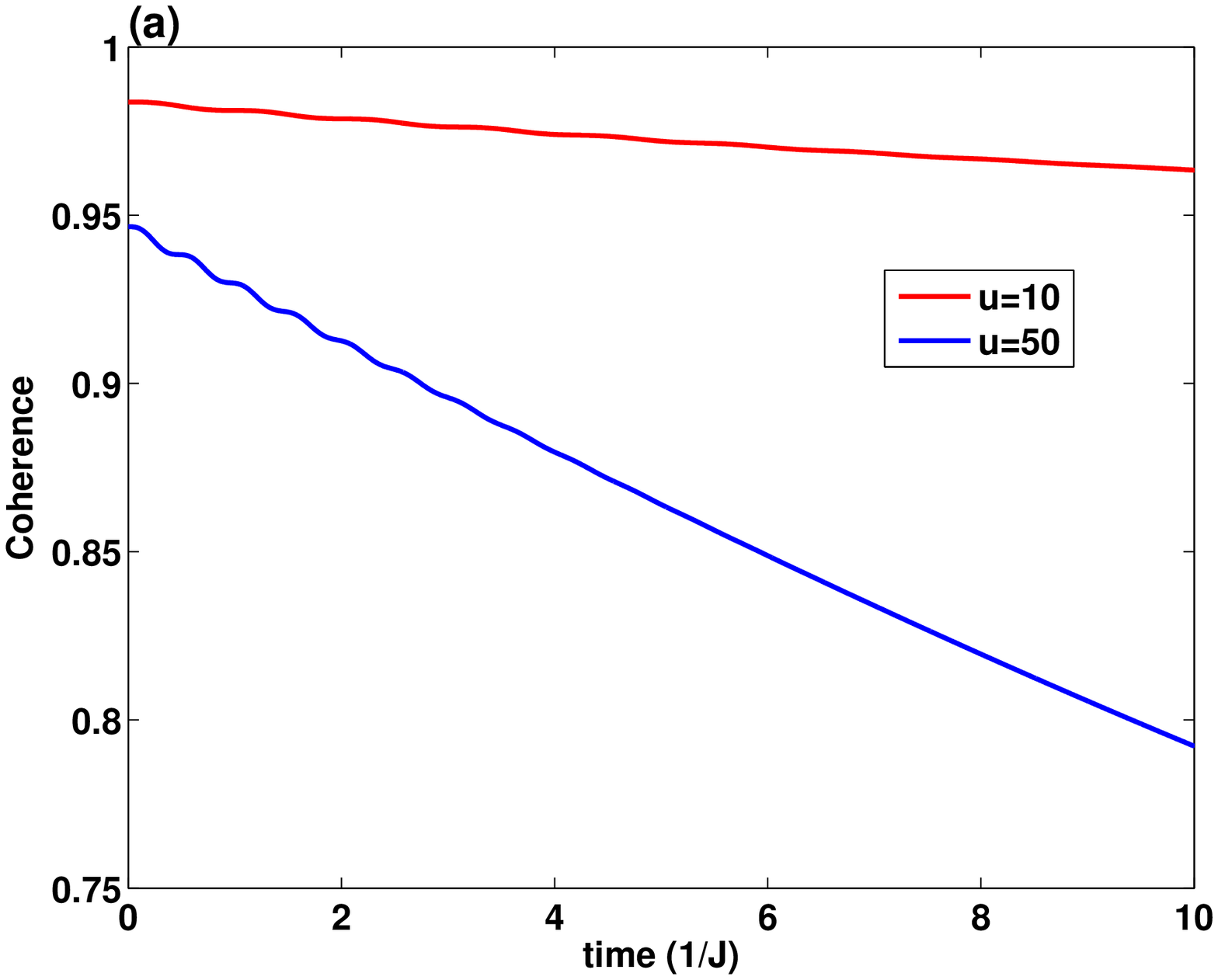} \\
      \includegraphics[width=\columnwidth]{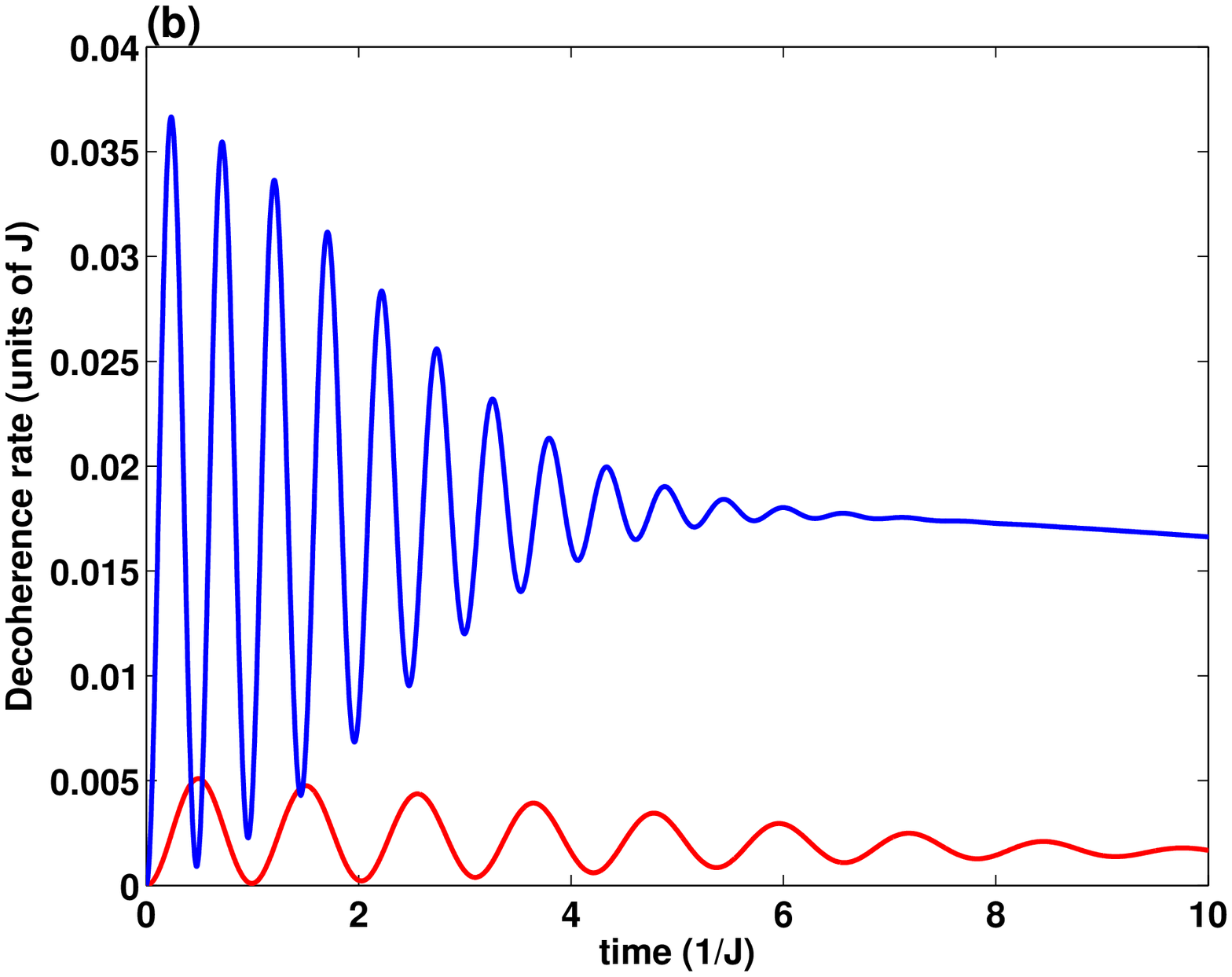}
\caption{%
Decoherence induced by loss. A loss rate $\gamma_{\rm loss}=0.08\ J$ is applied, such
that after $t=10/J$ only $\sim 45$\% of the atoms are left in the trap. During this time the coherence
drops due to interactions. We have used $N=50$ atoms in the numerical simulation, showing two curves, for
onsite interaction strengths $U = 0.2\,J$ [$u=10$] and $U = J$ [$u=50$]. (a) coherence as a function of time. (b) instantaneous rate of decoherence,
showing oscillations with half the Josephson period [$\omega_J = 3.32\,J$ and $7.14\,J$ for the two curves].
}
   \label{fig:gloss}
\end{figure}

In Fig.~\ref{fig:gloss} we show the time evolution of the coherence due to a loss process described by a dissipative
term as in Eq.~(\ref{eq:mastera}), which is driven by the annihilation operators $\hat{a}_L$ and $\hat{a}_R$
from both wells. For the parameters considered in Fig.~\ref{fig:gloss}, the decoherence rate is much smaller than the
loss rate, but it grows with the strength of the interaction $u$. The results of the numerical
simulation show that the initial rate of decoherence is close to zero [$c(0)=0$ in Eq.~(\ref{eq:gamma_loss})],
rather than to the lower value $\gamma_{\rm loss}
(1 - 1/\xi^2)/2N$.
However, the value of $s(t)$ is close to the expected value $\sin^2\omega_J t$.

For relatively strong interactions,
the decoherence rate may become
larger than the loss rate itself. This occurs when $\xi^4/2N
\sim u / 2 N > 1$, and may even happen in the Josephson
regime [see~Eq.~(\ref{ineq:Josephson})].

\section{Realizability and validity}
\label{sec:validity}

\begin{figure}[t!]
      \centering
\includegraphics[width=\columnwidth]{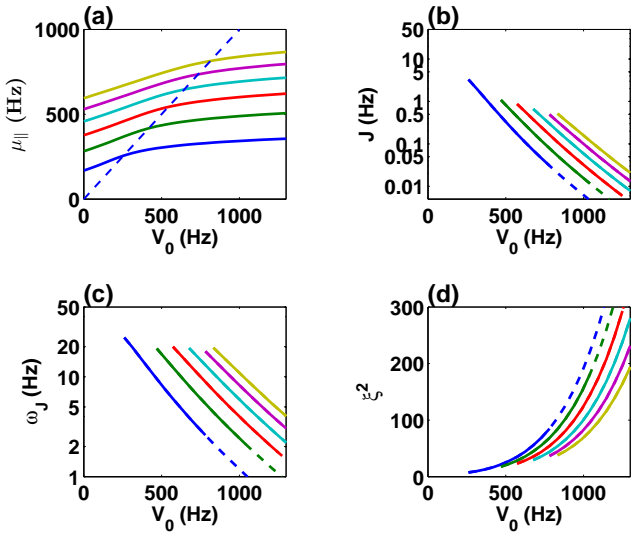}
\caption{Model validity range: the parameters of the Bose-Hubbard model for
a trapping potential having the form of Eq.~(\protect\ref{eq:Vxyz}) with parameters as in Fig.~\ref{fig:Elevels}
(transverse frequency $\omega_{\perp}=2\pi\times 500$\,Hz) and
the number of atoms $N=100,200,300,400,500,600$ [bottom to top in (a)]. (a) The longitudinal chemical potential $\mu_{\parallel}$
as a function of the barrier height $V_0$. We define
the validity range for each value of $N$ as the range where
$V_0>\mu_{\parallel}$ (to the right of the dashed line representing $\mu_{\parallel}=V_0$).
This validity criterion is presented in (b)-(d) by simply not drawing the curves where they do not adhere to this criterion.
(b) The tunneling matrix element $J$, (c) the Josephson frequency $\omega_J$, and
(d) the squeezing parameter $\xi^2$. The range where $\xi^2>N$ [dashed portions of the curves in (b)-(d)] is the Fock regime, where the coherence of the ground state drops to zero.
In this work we are interested in the
Josephson regime of Eq.~(\ref{ineq:Josephson}), where $\xi^2<N$.
When $\xi^2>2\sqrt{N}$, loss-induced decoherence is faster than the loss rate, as
happens for most of the curves in (d).}
\label{fig:validity500}
\end{figure}

The two-site Bose-Hubbard model, which was used in the previous sections, is fully valid (namely, applicable over the
whole phase space shown in Fig.~\ref{fig0}) only if the atom-atom interaction energy is much smaller than
the energy of excited modes in each well, such that higher-energy spatial modes are not
excited by the interaction.
For a single-well frequency $\omega$ this criterion implies
$gn\ll \hbar\omega$, where $n$ is the peak atomic density.
This requirement is equivalent to the healing length
$l_c=\hbar/\sqrt{mgn}$ being much longer than the trap width $L\sim \sqrt{\hbar/m\omega}$.
However, dynamics similar to that predicted by the two-mode model, such as Josephson oscillations and phase oscillations
of self-trapped populations, appear in mean-field (GP) calculations even if the interactions are much stronger,
as long as the atomic density changes during the evolution are small.
Since the GP approximation represents the classical limit of the two-site model, this suggests that the model
is valid for such dynamics, in which the populations in the two sites do not deviate much from their equilibrium
values and the spatial density can be derived from the two stationary modes $\phi_L$ and $\phi_R$.
Here we are interested in slow dynamics during which the population does not vary considerably, since the main change
is in the one-particle coherence.
With the aid of the mean-field approach described in Sec.~\ref{sec:system},
we therefore require that the dynamics involves only low-lying excitations of the two-mode system, whose energy is lower
than the energy splitting of higher spatial modes $E_j$ ($j\geq 2)$
[see Eq.~(\ref{eq:GP-excited-states})], such that only the pair $\phi_0$ and $\phi_1$ is populated.

In addition, the Hamiltonian [Eq.~(\ref{eq:Ham})] neglects interaction terms involving different modes, such as
a term proportional to $\hat{n}_L\hat{n}_R$~\cite{JaphaBand2011},
which may be significant if the two main modes $\phi_L$ and $\phi_R$ are not well separated in space.
We therefore define a simple validity criterion: that the barrier height $V_0$ is larger than the
longitudinal energy $\mu_{\parallel}$ of the condensate mode $\phi_0$.
With this condition we can show that the Josephson frequency $\omega_J$,
which defines the excitation energy in the two-mode system, is much smaller than the higher-mode excitation energy.
This ensures that, as long as low-energy states of the two-mode system are involved, higher spatial modes are
not excited and the system may still be described by the two modes.

\begin{figure}[t!]
      \centering
\includegraphics[width=\columnwidth]{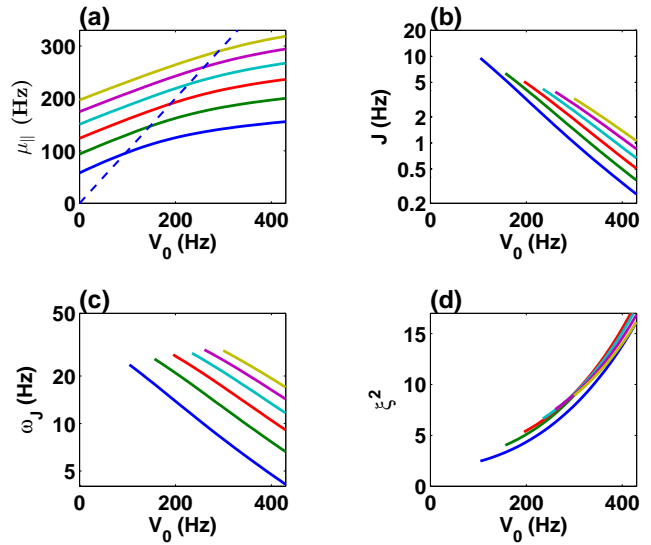}
\caption{Same as Fig.~\protect\ref{fig:validity500} with transverse frequency $\omega_{\perp}=2\pi\times 100
$\,Hz. Here $\xi^2<2\sqrt{N}$ for all curves, implying weak loss-induced decoherence over the parameter
range.}
\label{fig:validity100}
\end{figure}

In Figs.~\ref{fig:validity500} and~\ref{fig:validity100} we show the parameters of the Bose-Hubbard and
Josephson Hamiltonians in their validity range,
as calculated for $^{87}$Rb atoms in a potential having the form of Eq.~(\ref{eq:Vxyz})
with $d=5\,\mu$m and $\omega_x=2\pi\times 200$\,Hz, as in Fig.~\ref{fig:Elevels}.
The strength of the interaction is determined both by the total number of atoms $N$
and by the transverse frequency $\omega_{\perp}/2\pi$, which is equal to 500\,Hz in Fig.~\ref{fig:validity500}
and 100\,Hz in Fig.~\ref{fig:validity100}. In all cases the Josephson frequency is not larger than $30$\,Hz, which is smaller than
the excitation energy of higher modes, and thus ensures the validity of the model. However, for the
strongly interacting BEC in Fig.~\ref{fig:validity500} the squeezing factor $\xi^2$ may become larger than $N$, implying
that for such high barrier heights and low tunneling rates the system is in the Fock regime ($u>N^2$, represented
by the dashed curves). In this regime our model is not valid, since the coherence of the initial state
(the ground state) is already very low and the dynamics of coherence loss in general is dominated by phase diffusion.

Two differences between Fig.~\ref{fig:validity500} and Fig.~\ref{fig:validity100} are most noticeable. First,
the chemical potential of the BEC in tight confinement is much larger than that of the BEC in weaker
confinement, implying that the barrier height for which the two-mode model is valid is much lower for the latter. 
Second, the amount of squeezing is correspondingly much larger for the tightly confined BEC in Fig.~\ref{fig:validity500}. It follows that for a given rate $\gamma_{\rm loss}$
of atoms from the trap, we would expect a high decoherence rate for the tightly confined BEC,
since $\gamma_{\rm dec}/\gamma_{\rm loss}\sim \xi^4/4N>1$ for most of the parameter range
in Fig.~\ref{fig:validity500}. Conversely, we should expect very low decoherence rates relative to the loss rate
in the case of weak confinement as in Fig.~\ref{fig:validity100}.

In the next section we present some experimental results that give realistic loss rates near
an atom chip. This will enable us to discuss expected decoherence rates for atomic
Josephson junctions in similar scenarios.


\section{Measurements of atom loss near an atom chip}
\label{sec:lifetime}

\begin{figure}[t!]
   \centering
   \includegraphics[width=0.5\textwidth]{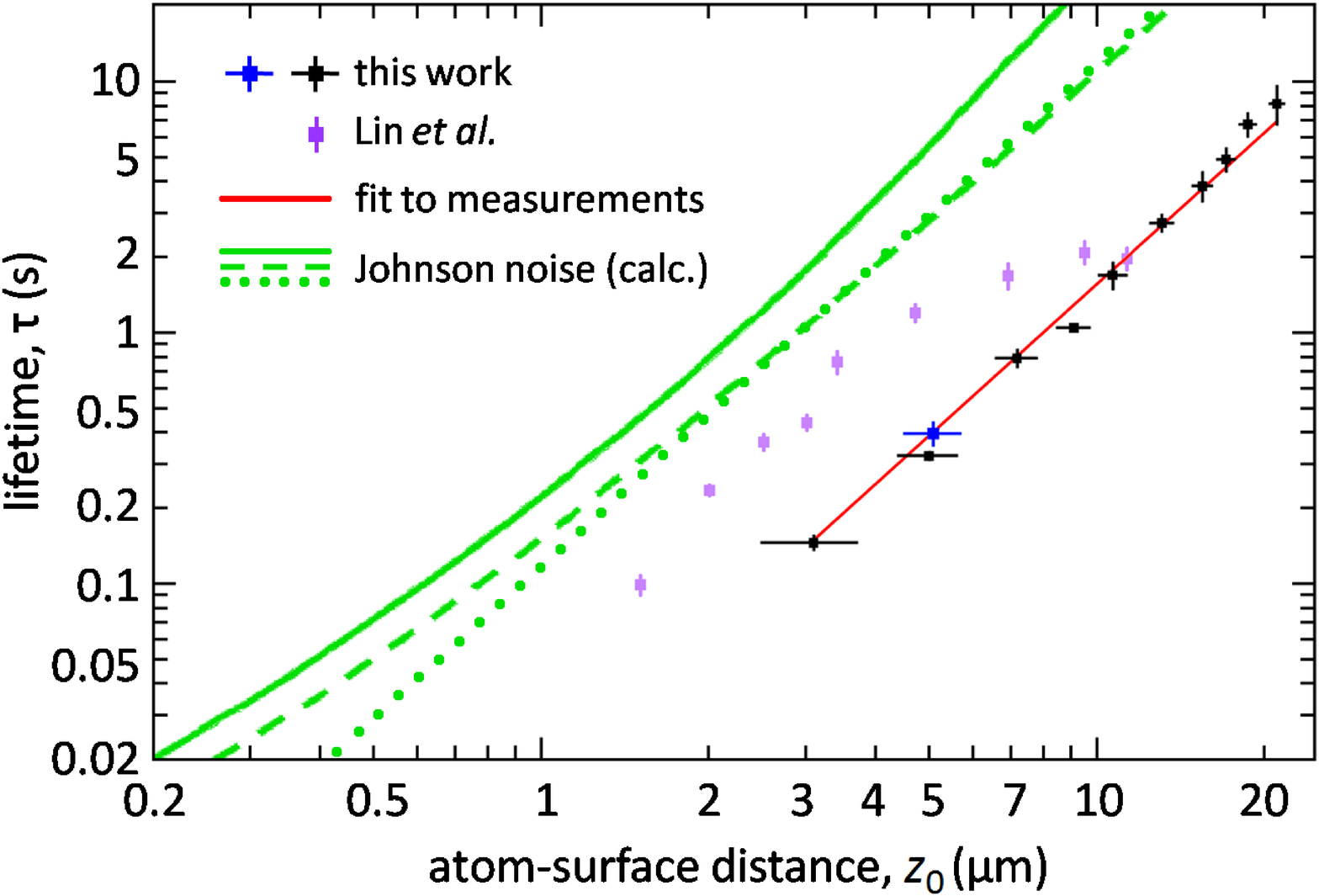}
   \caption{Atom chip noise: lifetimes measured for a range of atom-surface distances. The black data points are for thermal $^{87}$Rb atoms, while the blue datum is for a~BEC. For comparison, we show data for conditions similar to those of our measurements and over a similar range of distances, but using no current in the surface closest to the atoms and therefore exhibiting much weaker technical noise\,~\cite{PRL92-050404}). The solid red line is the best-fit line for an assumed quadratic dependence using our measurements for thermal atoms. The green curves are calculated for spin flips due to Johnson noise only. The solid green curve also accounts for cascading and
the lateral wire geometry~\cite{EurPhysJD35-87}; the dashed
green curve is valid also at distances comparable to the gold
layer thickness but does not include cascading and the wire geometry;
the dotted green line is the simple power law of
Eq.~(\ref{eq:Johnson}).}
   \label{fig:lifetime}
\end{figure}

We investigate the expected dephasing due to the nearby surface by measuring the lifetime of atomic clouds held at varying atom-surface distances $z_0$.
The cloud temperature is about $\rm2\,\muK$; distances are measured by reflection imaging~\cite{Zhou2014}.
Our data, adjusted for the lifetime due to vacuum of about~$\rm30\,s$, are shown in Fig.~\ref{fig:lifetime}.

For comparison with theory, let us for the moment ignore cascading effects (\ie\ transitions
$|2,2\rangle \longleftrightarrow |2,1\rangle \longrightarrow |2,0\rangle$
amongst Zeeman sub-levels that can re-populate the initial state),
and the finite lateral extent of the current-carrying wire  (\ie\  we initially assume a layer of infinite extent).
We assume that the measured lifetimes $\tau_{\rm meas}$ are due to Johnson and technical noise, and we write~\cite{PRA72-042901,EurPhysJD51-173}:
\begin{eqnarray}
\tau_{\rm meas}^{-1} &=&
\tau_{\rm Johnson}^{-1}+\tau_{\rm tech}^{-1}
	\label{eq:measured}
	\\[6pt]
\frac{ 1 }{ \tau_{\rm Johnson} } &=&
\frac{ c_1 }{ z_0^2 }
= \left(\frac{3}{8}\right)^2 \frac{\bar{n}_{\rm th}+1}{\tau_0}	    \left(\frac{c}{\omega}\right)^3 \frac{2h}{\delta^2 z_0^2}
	\label{eq:Johnson}
	\\[6pt]
\frac{ 1 }{ \tau_{\rm tech} } &=&
\frac{ c_2 }{ z_0^2 }
	= \left(\frac{\mu_0\mu_F}{2\pi\hbar}\right)^2
	\frac{ {\mathcal I}^2 }{ z_0^2 }\ ,
	\label{eq:tech}
\end{eqnarray}
where $\bar{n}_{\rm th} = k_BT/\hbar\omega \approx 2\times10^7$, 
and $\tau_0$ is
the free-space lifetime at the Larmor
frequency $\omega$~\cite{nbar}.
Our experimental parameters are: gold wire with thickness
$h=\rm0.5\,\mum$; $\omega/2\pi=500\rm\,kHz$; and $T=\rm400\,K$ due
to Joule heating of the atom chip wire.
The skin depth $\delta \approx \rm130\,\mum$ is calculated with
a temperature-dependent resistivity $\rho( T ) = \varrho T$
for gold~\cite{NIST}. With these substitutions, the Johnson
lifetime is actually independent of $\omega$ and $T$. The
scaling with distance is valid in the intermediate regime
$h \ll z_0 \ll \delta$~\cite{PRA72-042901}. The lifetime due
to technical noise involves the current noise spectrum ${\cal I}$
(in ${\rm A} / \sqrt{\rm Hz}$)~\cite{EurPhysJD51-173};
it scales with the same power of the distance $z_0$.

The experimental data of Fig.~\ref{fig:lifetime} are fitted
on a logarithmic scale and conform to the exponent
$\tau_{\rm meas} \propto z_0^2$ very well. The fit yields
$c_1 + c_2 \approx 65\,{\rm \mu m^2/s}$.
From Eq.~(\ref{eq:Johnson}), we estimate the coefficient~$c_1$
for Johnson noise, finding $c_1 \approx 8.5\,\rm \mum^2/s$.
This yields in turn~$c_2\approx56\,\rm \mum^2/s$ and
corresponds
to lifetimes for Johnson and technical noise
of~$\rm3\,s$ and~$\rm0.4\,s$, respectively at~$z_0=\rm5\,\mum$.
Re-introducing cascading and geometrical effects\,~\cite{EurPhysJD35-87} doubles the estimated lifetime to~$\rm6\,s$ for the Johnson noise component in these atom chip measurements
(thick green lines in Fig.~\ref{fig:lifetime}, see caption for
details). The power law $\sim z_0^2$ for the lifetime
becomes invalid only for distances $z_0 < 2\,\rm\mum$, becoming comparable to the gold wire thickness.

We conclude that in our experiment, technical noise is the dominant
cause of loss, by at least a factor of~$6$.
We note that both the geometrical and cascading factors lengthen the calculated lifetime due to Johnson noise, so the conclusion that technical noise dominates our losses is reinforced. This may also explain
the difference in lifetime compared to Ref.~\cite{PRL92-050404}.
From Eq.~(\ref{eq:tech}) we obtain a current noise spectral density of~${\mathcal I}=0.9\rm\,nA/\sqrt{Hz}$. This is much
larger than shot noise, by at least one order of magnitude,
strengthening the hypothesis of a technical origin for the
current noise.


\section{Discussion}
\label{sec:discussion}

In this work we have investigated the effect of noise on the decoherence of a BEC, initially in its ground state in a double-well
potential. Specifically, we have discussed three dephasing mechanisms:
(a) direct dephasing from short correlation length (Johnson) noise or asymmetric (technical or Johnson) noise. This can be represented by random rotations about the $S_3$-axis of the Bloch sphere. The phase-space distribution in Fig.~\ref{fig0}
then diffuses in the $\varphi$-direction so that the relative phase between the left and right wells gets randomized (phase noise);
(b) dephasing due to population difference fluctuations (number noise). 
On the Bloch sphere, these correspond to random rotations around the $S_2$-axis. 
They may originate from an overlap between the spatial modes of atoms in the two sites and 
fluctuating magnetic fields with a short correlation length. 
While Fig.~\ref{fig1} shows that rotation about the $S_2$-axis gives rise to an enhanced decoherence rate, 
the physical source of these rotations is weak and its contribution may be assumed to remain small even when enhanced;
(c) dephasing due to losses induced by both Johnson and technical noise. 
This decoherence process involves an enhancement effect similar to that of number noise, which may make it dominant
for strong atom-atom interactions.

Of the three dephasing mechanisms examined, we find that two are dominant: direct dephasing~(a) and loss-induced decoherence~(c). Direct dephasing can be suppressed by atom-atom interactions (up to a factor of~2).  However, interactions play quite a different role in the context of loss-induced decoherence. While loss induced by noise has no effect on the coherence of the remaining atoms if the initial state is a coherent state of non-interacting atoms,
it may produce considerable decoherence if atom-atom interactions are dominant. Such enhanced decoherence would appear for any kind of noise which tends to change (randomize) the relative number of particles in the two wells. 

It is worthwhile to note a general result following from our derivation. Although we have shown that symmetric noise with a long correlation length does not lead to direct dephasing
of the type (a) above if the density matrix of the system is diagonal in the basis of states with a well-defined total number $N$, 
a master equation with a stochastic term as in Eq.~(\ref{eq:Lrho-const}) may still lead to the decay of off-diagonal density matrix elements. 
If one would adopt a symmetry-breaking approach to BEC~\cite{Leggett1991}, where
the density matrix involves superpositions between different number states, these would decay under this 
interaction. This scenario would provide a typical example of a dynamically emerging ``superselection rule'' stating that superpositions of different atom numbers are forbidden.

Our results are well established, as long as the two-mode model is valid, since their main 
characteristics are derived independently
by three different approaches: semiclassical phase-space methods, analytical calculations using a
linear excitation approximation, and exact numerical calculations for relatively small atom numbers.
However, we note two limitations of our model. First, it does not take into account possible effects of heating or cooling of the Bose gas
due to transitions between the two main modes and higher-energy spatial modes in the trap.
Such transitions may be driven by components of the external noise which have correlation lengths on the order
of the single trap width, or by external magnetic fields which cause deformations of the potential with higher-order
spatial dependence. Such processes may affect the dynamics of decoherence in the double-well trap
but are beyond the scope of this paper. 
Second, our model is valid for a BEC with a macroscopic number of 
atoms, while dephasing in a system of a few atoms, which may be relevant to atomic circuits, would have to be 
treated separately. 

Finally, the results of this work allow an estimation of the accessible range of parameters for an atomic Josephson junction
permitting operation over a reasonable duration of time without significant decoherence. As we have shown in
Sec.~\ref{sec:lifetime}, typical values of the Johnson noise at a distance of 5\,$\mu$m from the surface
cause losses at a rate of less than
0.5\,s$^{-1}$. At this distance, the rate of dephasing due to Johnson noise is expected to be on the same order
as the loss rate and we therefore expect that such dephasing will enable coherent operation for a time scale of a few seconds.
This time scale could even be doubled in the presence of squeezing due to interactions,
as predicted by our theory in Sec.~\ref{sec:main_A}. The main source of noise in our experiments is found to be technical
noise, which is not expected to directly cause dephasing due to its long correlation length, provided it does not contain strong
asymmetric components. Technical noise, however, induces loss, and decoherence due to loss is expected to be significant if the BEC is strongly interacting due to tight transverse confinement as in Fig.~\ref{fig:validity500}. In this case we expect that the squeezing factor is so large that the coherence time is shorter than the trapping lifetime. However, in the case of weak confinement, as in Fig.~\ref{fig:validity100},
we expect that the decoherence rate due to loss is smaller by a factor of $\xi^4/4N\lesssim 4/N$, which is much
smaller than the loss rate. 

To conclude, at distances of a few $\rm\mu m$ (for which accurately controllable tunneling barriers may be formed) and within the framework of a total spatial decoherence rate equal or smaller than the loss rate, we find that tunneling rates of about $0.1-10$ Hz, or Josephson oscillation frequencies $\omega_J/2\pi$ of about $2-25$ Hz, may be obtained (Fig.~\ref{fig:validity100}), depending on the barrier height, degree of transverse confinement, and number of atoms. This provides a wide dynamic range in the operation of a tunneling barrier for atomtronics.

\acknowledgments

We thank David Groswasser for support with the experimental system. This work is funded in part by the Israeli Science Foundation (1381/13), the European Commission ``MatterWave'' consortium (FP7-ICT-601180), and the German DFG through the DIP program (FO 703/2-1). We also acknowledge support from the~PBC program for outstanding postdoctoral researchers of the Israeli Council for Higher Education. MK acknowledges support from the Ministry of Immigrant Absorption (Israel) and AV acknowledges support from the Israel Science Foundation (346/11).

\vfill\eject

\bibliography{bec-dephasing}

\begin{thebibliography}{10}
\bibitem{nbar}
Ref.~\cite{PRA72-042901} gives $\tau_0=3\times10^{25}\rm\,s$
  at~$\omega=\rm400\,kHz$. Note that the discussion following Eq.~(4) of
  Ref.~\cite{PRA72-042901} implies a value of $\bar{n}_{\rm
  th}=0.75\times10^7$.
\end{thebibliography}

\end{document}